\documentclass[aoas,MSNbibl,seceqn,nameyear,rotating,dvips]{arximspdf}
\usepackage{multirow}
\usepackage{dcolumn}
\usepackage{graphicx}

\doi{10.1214/11-AOAS502}
\volume{6}
\issue{1}
\pubyear{2012}
\firstpage{334}
\lastpage{355}

\makeatletter

\def\bsuffix #1{#1}

\newcolumntype{d}[1]{D{.}{.}{#1}}

\renewcommand{\bm}[1]{\bolds{#1}}
\newcommand{\bmm}[1]{\mathbf{#1}}

\makeatother

\begin{document}
\begin{frontmatter}

\title{Bayesian joint modeling of multiple gene networks and diverse genomic data to identify target genes of a transcription factor\thanksref{T1}}
\runtitle{Joint modeling of networks and genomic data}

\thankstext{T1}{Supported in part by NIH Grant HL65462.}

\begin{aug}
\author[A]{\fnms{Peng} \snm{Wei}\corref{}\thanksref{T2}\ead[label=e1]{Peng.Wei@uth.tmc.edu}}
\and
\author[B]{\fnms{Wei} \snm{Pan}\ead[label=e2]{weip@biostat.umn.edu}}
\runauthor{P. Wei and W. Pan}
\affiliation{University of Texas School of Public Health
and University of Minnesota}
\address[A]{Division of Biostatistics\\
\quad and Human Genetics Center\\
Univeristy of Texas School of Public Health\\
1200 Herman Pressler Dr, RAS E817\\
Houston, Texas 77030\\
USA\\
\printead{e1}}
\address[B]{Division of Biostatistics\\
School of Public Health \\
University of Minnesota \\
Minneapolis, Minnesota 55455\\
USA\\
\printead{e2}} 
\end{aug}

\thankstext{T2}{Supported in part by a start-up fund from
the University of Texas School of Public Health.}

\received{\smonth{8} \syear{2010}}
\revised{\smonth{6} \syear{2011}}

%
\begin{abstract}
We consider integrative modeling of multiple gene networks and diverse
genomic data, including protein-DNA binding, gene expression and DNA
sequence data, to accurately identify the regulatory target genes of a
transcription factor (TF). Rather than treating all the genes equally
and independently a priori in existing joint modeling approaches,
we incorporate the biological prior knowledge that neighboring genes on
a gene network tend to be (or not to be) regulated together by a TF. A
key contribution of our work is that, to maximize the use of all
existing biological knowledge, we allow incorporation of multiple gene
networks into joint modeling of genomic data by introducing a mixture
model based on the use of multiple Markov random fields (MRFs). Another
important contribution of our work is to allow different genomic data
to be correlated and to examine the validity and effect of the
independence assumption as adopted in existing methods. Due to a fully
Bayesian approach, inference about model parameters can be carried out
based on MCMC samples. Application to an E. coli data set,
together with simulation studies, demonstrates the utility and
statistical efficiency gains with the proposed joint model.
\end{abstract}

%

\end{frontmatter}

\section{Introduction}

In this paper we consider integrative modeling of multiple sources of
genomic data and gene networks to accurately identify the regulatory
target genes of a transcription factor (TF). TFs, a class of regulatory
proteins, play a central role in controlling gene expression: a TF
stimulates or inhibits its target gene's transcription into messenger
RNA (mRNA) by binding to some specific DNA subsequences in the gene's
promoter region. In our motivating example, we are interested in
identifying the target genes of LexA in E. coli. LexA is an
important TF involved in DNA repair and cell division: it is a
repressor for genes involved in the ``SOS'' response whose transcription
is induced in response to DNA damage due to ultraviolet (UV) or
chemical exposures [\citet{ZhaPigRic10}]. Under normal growth
conditions, LexA binds to the promoter regions of these ``SOS'' genes,
repressing their transcription. When DNA becomes extensively damaged,
the LexA repressor is cleaved and loses its function. As a result, the
expression of ``SOS'' genes is induced, and DNA repair ability in the
cells is enhanced. Recently, LexA was shown to be essential in the
acquisition of bacterial mutations which lead to resistance to some
antibiotic drugs [\citet{Ciretal05}]. Therefore, a thorough
understanding of LexA regulation is not only crucial to the elucidation
of the DNA repair mechanism in E. coli, a common model
microorganism, but also beneficial to antibiotic drug development
[\citet{ButZfuBus09}].

The task of identifying the target genes of a TF can be approached by
using ChIP-chip data (also called DNA-protein binding data or
genome-wide location analysis), which provide evidence about
genome-wide physical binding sites of a specific TF in living cells.
However, those DNA--TF interactions may not be functional in terms of
regulating gene expression because other conditions such as binding of
co-regulators and recruitment of RNA polymerase II complex are also
needed to initiate the target gene's transcription. Two other types of
genomic data, also available for LexA, provide complementary
information about TF-gene regulation: microarray gene expression data
comparing expression changes before and after knocking-out or mutating
a TF-coding gene, and DNA sequence data which are aligned and scanned
to find specific binding sites of a TF, called consensus sequence or
motif. Although extremely valuable, these two data sources provide only
partial information: for expression data, genes that are directly or
indirectly regulated by the TF will all show changes in expression
levels, while DNA sequence data provide only potential binding sites
which may or may not eventually be bound by the TF. Because each data
source measures different aspects of TF-gene regulation, and
high-throughput data are inherently associated with high noise levels,
using one type of data alone may result in high false positives or
false negatives.

In contrast, it is now widely recognized that an integrative analysis
of multiple types of genomic data should be more efficient in
identifying the target genes of a TF [see \citet{Wanetal05},
\citet{JenCheSto07}, \citet{PanWeiKho08},
\citet{Xieetal10} and references therein]. There are two main
classes of joint modeling approaches in the literature: regression
methods and mixture model methods. First, in a regression framework,
one type of data (e.g., ChIP-chip binding data or DNA sequence data) is
regressed on another type of data [e.g., gene expression data;
\citet{Conetal03}, \citet{SunCarZha06}, \citet{WeiPan08N2}]. In
particular, \citet{JenCheSto07} proposed a Bayesian regression
model in a variable selection framework to combine all three sources of
data to construct gene regulatory networks (i.e., a set of multiple TFs
and their regulatory target genes). Note that regression-based methods
require a large number of replicated expression arrays, which are not
applicable to the LexA data to be analyzed here. Second, in a mixture
model framework, inference is based on the posterior probability of
being a target given gene-specific measurements of different sources of
data. \citet{Wanetal05} proposed a parametric mixture model for
both DNA sequence data and expression/binding data;
\citet{PanWeiKho08} extended the mixture model of Wang et al. to
one that is able to integrate all three data sources to detect the
targets of a TF. Conditional independence is commonly assumed in a
mixture joint model, that is, different sources of data are independent
given that a gene is or is not a target, which may or may not hold in
practice. In particular, it has been reported in the experimental
biology literature that the binding strength of LexA to its target
genes depends on the extent to which the binding site matches the
canonical motif of LexA [\citet{Mic05}, \citet{ButZfuBus09}].
Hence, the conditional independence assumption seems incorrect, at
least for the binding and sequence data, motivating us here to extend
the parametric mixture model of Pan et al. to allow conditional
dependence. We propose to summarize each data source with a scalar
summary statistic for each gene, and, thus, the three sources of
genomic data can be conveniently modeled by a trivariate normal mixture
model. Moreover, by adopting a fully Bayesian approach, we are able to
make inference about the conditional correlation structures for all
three data sources based on Markov chain Monte Carlo (MCMC) samples.

In addition to relaxing the conditional independence assumption,
another key contribution of our proposed method here is to allow
incorporation of multiple gene networks into joint modeling of diverse
types of genomic data to detect the targets of a TF. Gene networks
represented by undirected graphs with genes as nodes and gene--gene
interactions as edges provide a powerful means to concisely summarize
biological knowledge that is accumulated over thousands of experiments.
An emerging class of statistical methods is to incorporate gene network
information into analysis of genomic data [Wei and Li
(\citeyear{WeiLi07}, \citeyear{WeiLi08}), \citet{LiLi08}, Wei and
Pan (\citeyear{WeiPan08N1}, \citeyear{WeiPan10})]. In particular,
\citet{WeiLi07} proposed a discrete Markov random field
(MRF)-based mixture model to incorporate gene network information into
statistical analysis of gene expression data to boost the power for
detection of differentially expressed genes. \citet{WeiPan10}
proposed a Bayesian implementation of the MRF-based mixture model of
\citet{WeiLi07}, and compared it with the Gaussian MRF-based
mixture model of \citet{WeiPan08N1}. The network-based methods are
motivated by the biological fact that neighboring genes on a network,
for example, co-expression or functional coupling gene network, are
more likely to be co-regulated by a TF than nonneighboring ones.

\begin{figure}
\begin{tabular}{@{}cc@{}}

\includegraphics{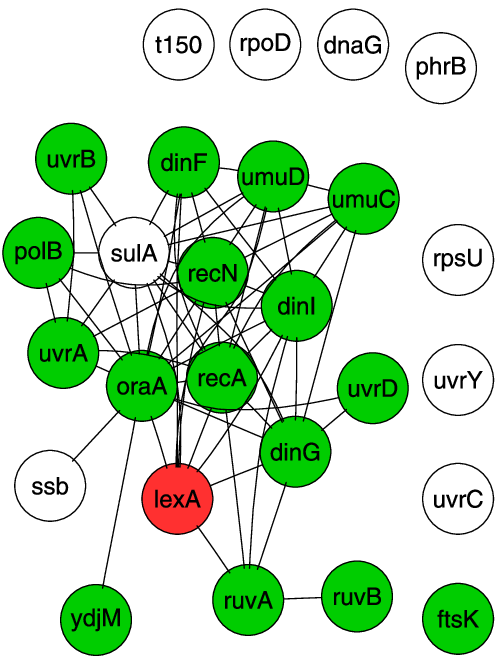}
 & \includegraphics{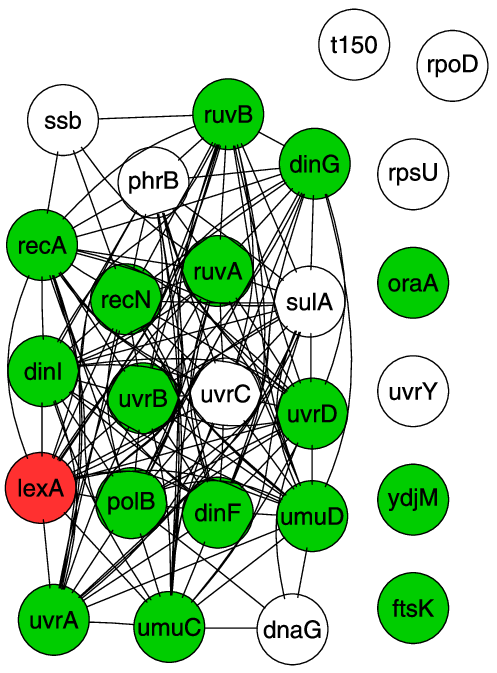} \\
(a) & (b)
\end{tabular}
\caption{Subnetworks, one from each of the following two networks,
consisting of LexA's known (colored/shaded nodes) and putative (blank
nodes) target genes as available from RegulonDB. The two gene networks
are: \textup{(a)} co-expression network, and \textup{(b)}
GO-induced functional coupling network.}\label{jointF1}
\end{figure}

One limitation of existing network-based methods, including the
aforementioned ones, is that only a single gene network is allowed to
be integrated with a single type of genomic data. However, as
biological knowledge accumulates rapidly, multiple gene networks become
available. For humans, existing gene networks include the KEGG gene
regulatory network [\citet{KanGot02}], the functional gene
network of \citet{Fraetal06} and several protein--protein
interaction (PPI) networks, for example, the Human Protein Reference
Database (HPRD) of \citet{Praetal09} and the Online
Predicted Human Interaction Database (OPHID) of \citet{BroJur05}, among others. Interactions between two genes in different
networks may have different biological implications. For example, for
E. coli two gene networks can be used to analyze the LexA data:
(1)~a co-expression network constructed based on a compendium of gene
expression microarrays, where two genes are direct neighbors if their
expression levels were highly correlated across about 400 experimental
conditions; (2) a functional coupling network induced by a Gene
Ontology [GO; Ashburner et al. (\citeyear{Ashetal00})] semantic similarity, where
two genes are direct neighbors if their functional annotations are
specific and close enough in the GO, a database containing the most
comprehensive existing knowledge of gene function. Figure \ref
{jointF1} shows subnetworks, one from each of the aforementioned
networks, consisting of LexA's known and putative target genes as
%
available from RegulonDB [\citet{Gametal08}], a~database
containing all known TF-gene regulatory interactions in E. coli.
As we can see, a~gene may have different sets of direct neighbors
according to different networks. This is in part because edges in
different networks reflect different perspectives of gene--gene
interactions, for example, co-expression or co-function, and in part
because of incomplete or simply wrong annotation shown by a network.
Since the two gene networks contain partial yet complementary
information about gene--gene interactions, integrating both of them with
ChIP-chip binding, gene expression and DNA sequence data is expected to
boost the power for detecting the target genes of LexA. As a key
contribution, here we propose a~mixture model to address this problem
based on the use of multiple MRFs. Statistical inference is carried out
in a fully Bayesian framework. The proposed method can be easily
extended to integrate more gene networks and more types of genomic
data, providing a general statistical framework for integrative
analysis of genomic data.

The rest of this article is organized as follows. We first describe the
LexA data including ChIP-chip binding, gene expression, DNA sequence
data and two gene networks for E. coli. Next, we introduce a
multivariate normal mixture model for joint modeling of multiple
sources of genomic data only, followed by a unified mixture model for
integrating multiple gene networks and genomic data based on the use of
multiple MRFs. We discuss statistical inference for the proposed models
in a fully Bayesian framework. Parameter estimates are based on MCMC
samples. We apply the new methods to the LexA data to identify its
regulatory target genes. We evaluate the proposed methods' predictive
performance by comparing the results with the known and putative
targets listed in RegulonDB (v6.4). We also show results from
simulation studies to investigate the conditional independence
assumption as well as the effects of integrating multiple networks and
diverse types of genomic data. We end with a discussion of some
existing issues and possible future work.

\section{The data}
\label{chap4data}
\subsection{ChIP-chip binding, gene expression and DNA sequence data}
\label{chap4genomicdata} The ChIP-chip binding data, gene expression
data and DNA sequence data were extracted and processed from three
sources as reported in \citet{Wadetal05}, \citet{Couetal01}
and RegulonDB (v4.0), respectively.

The ChIP-chip data included two LexA samples (called LexA$_1$ and
LexA$_2$, resp.) and
two control samples [one Gal4 and one MelR (no Ab, no antibody)
samples] hybridized on four Affymetrix Antisense Genome Arrays, respectively.
First, the arrays were background corrected with the MAS~5 algorithm,
followed by quantile normalization. Second, four $\log_2$ intensity
ratios (LIRs) were calculated, corresponding to the four combinations
of any two arrays, for each probe: LexA$_1$/Gal4, LexA$_1$/no Ab,
LexA$_2$/Gal4, LexA$_2$/no Ab; a large LIR indicated a locus containing
enriched LexA, that is, a binding site of LexA. Third, for each of the
four array combinations, the LIRs were smoothed over all probes with a
sliding window of 1,250 base pairs~(bp) along the chromosome. Finally,
gene $i$'s binding score $B_i$, a summary statistic measuring the
relative abundance of the TF binding to the gene, was taken to be the
average of its four LIR peaks from its coding region, or if there were
probes from its intergenic region, $B_i$ was the larger of (i) the
average of its four LIR peaks from its coding region and (ii) that from
its intergenic region.

The expression data were drawn from four cDNA microarrays profiling
gene expression levels for the wild type before and 20 minutes after UV
treatment, and for the LexA mutant before and 20 minutes after UV
treatment; a common control sample was used for each array. Two-channel
intensities on each array were normalized using the loess local
smoother to eliminate dye bias, as implemented in the R package
\texttt{sma} [\citet{YanDud02}].
Suppose that normalized log-ratios of the two-channel intensities for
gene $i$ on the four arrays were $M_{1i},\ldots,M_{4i}$, respectively, then
the summary statistic for gene expression data was taken as $E_i
=(M_{2i}-M_{1i})-(M_{4i}-M_{3i})$.
Because LexA is known to be a repressor of some ``SOS'' response genes,
it is expected that the regulatory targets of LexA should have
larger values of~$E_i$'s (i.e., expression changes).

The DNA sequence data were obtained as following. Ten known binding
sites of LexA were downloaded from RegulonDB (v4.0), involving nine
genes each with one binding site and gene LexA with two binding sites.
These ten binding sites were input into MEME [\citet{BaiElk95}]
to find a top consensus sequence (motif). scanACE [\citet{Rotetal98}] was then used to scan the whole genome with a very low
threshold such that at least
one subsequence matching the motif could be obtained for most genes;
the maximum of all the matching scores for gene $i$ was taken as $S_i$,
the summary statistic for the sequence data.

\begin{table}
\tablewidth=260pt
\caption{Some data from the LexA data set}\label{tabBES}
\begin{tabular*}{\tablewidth}{@{\extracolsep{\fill}} l d{2.3} d{2.3} d{2.3}@{}}
\hline
\textbf{Index} & \multicolumn{1}{c}{\textbf{Binding} $\bolds{(B_{i})}$}
& \multicolumn{1}{c}{\textbf{Expression} $\bolds{(E_{i})}$} &
\multicolumn{1}{c@{}}{\textbf{Sequence}
$\bolds{(S_{i})}$} \\
\hline
GENE1 & -0.490 & 0.076 & 15.573 \\
GENE2 & 2.275 & 2.777 & 23.968 \\
GENE3 & 0.619 & 1.377 & 24.164 \\
GENE4 & 0.210 & -0.208 & 15.464 \\
GENE5 & 0.120 & -0.346 & 13.055 \\
\hline
\end{tabular*}
\end{table}

After combining the three data sources and deleting genes with any
missing values, we obtained $G=3\mbox{,}779$ genes in the combined data. Table
\ref{tabBES} shows a small portion (5 of 3,779 genes) of the resulting
data set.

\subsection{Gene networks for E. coli}
Two gene networks were constructed for E.~coli as mentioned
before: a co-expression network and a functional coupling network.

The co-expression gene network was derived from 380 microarray
experiments across a variety of conditions, available at the Many
Microbe Microarrays Database [M3D; \citet{Faietal}]. Two
genes were direct neighbors if the Pearson correlation coefficient of
their expression profiles across the~380 experiments was greater than
0.65, resulting in a network with 3,208 nodes (genes) and 86,791 edges
(interactions). The cutoff 0.65 was chosen so that the resulting
network was neither too dense, including many false positive
interactions, nor too sparse, failing to include many true positive
interactions. As a comparison, a cutoff of 0.6 would lead to 147,563
interactions, while a cutoff of 0.7 would result in 46,666
interactions. We also performed sensitivity analysis to investigate how
robust the network-based analysis results are to different cutoffs for
the co-expression network (see Section \ref{sec43} for details).

\begin{figure}

\includegraphics{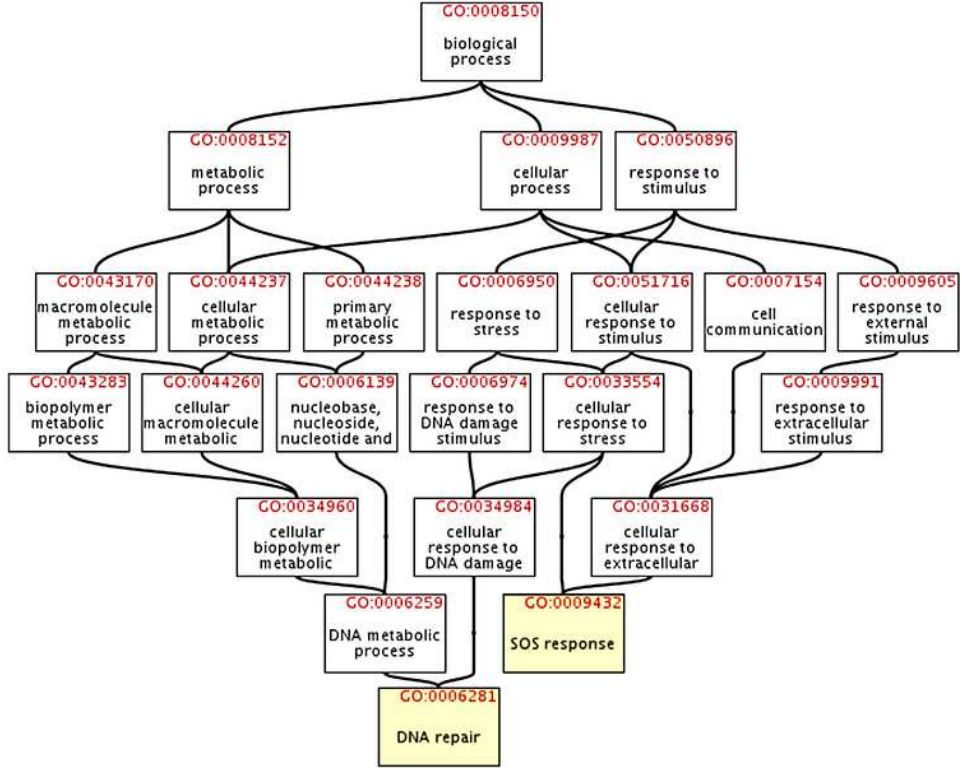}

\caption{The combined directed acyclic graph (DAG) of DAGs induced from
the GO terms ``DNA repair'' (\textit{GO:0006281}) and ``SOS response''
(\textit{GO:0009432}). lexA and dinG, two known target genes of TF LexA, are
annotated in both terms. Because there are 6 and~5 nodes in the longest
paths from ``DNA repair'' and ``SOS response'' to the root node
``biological process,'' respectively (the root node itself is not
counted), the GO similarity between lexA and dinG is 6. The graph was
adapted from QuickGO GO Browser (\protect\texttt{\href{http://www.ebi.ac.uk/QuickGO/}{http://}
\href{http://www.ebi.ac.uk/QuickGO/}{www.ebi.ac.uk/QuickGO/}}).}
\label{jointF2}
\end{figure}

The functional coupling gene network was induced from the Gene Ontology
(GO), a compendium of existing knowledge, derived from various sources,
about gene function. GO is structured as a directed acyclic graph
(DAG): each node corresponds to a GO category; a parent node represents
a more general biological function, whereas its child node is a
subclass or a part of it; any gene in a child node is necessarily in
its parent node. For example, GO category GO:0033554 with annotation
``cellular response to stress'' has a child node GO:0009432 with a more
specific annotation ``SOS response.'' The GO similarity between two
genes is defined as the maximum number of common nodes in all paths
back to the root node of the ontology (``biological process'') from all
nodes to which those genes are assigned [see \citet{Wuetal05} for
more details]. If the GO similarity between two genes is large, then at
a very specific level the two genes are involved in at least one common
biological process. Figure \ref{jointF2} illustrates a DAG induced
from the GO. We computed the GO similarity for each pair of genes. Two
genes were direct neighbors on the induced functional coupling network
if their GO similarity was no less than five, which means there were at
least five common nodes in their shared longest path back to the root
node ``biological process'' from all nodes in which they are annotated.
Figure \ref{jointF2} shows an example of how to calculate the GO
similarity between two genes. The induced network has 1,644 nodes and
116,422 edges.

Some summary statistics and sample subnetworks of the two gene networks
can be found in Table \ref{tabnet} and Figure \ref{jointF1},
respectively. The networks differ substantially in the density of edges
due to different definitions of gene--gene interactions.

\section{Statistical methods}
\subsection{Notation}
Our goal is to identify regulatory target genes of a given TF based on
given ChIP-chip binding, gene expression and DNA sequence data. We
assume that the three data sources have been summarized as~$(B_{i},\allowbreak E_{i},S_{i})$
for each gene $i$, for $i=1,\ldots,G$, as
described in Section \ref{chap4genomicdata}. Depending on the latent
(unobserved) state of gene $i$, that is, whether it is a~target or not,
we have $T_{i}=1$ or $T_{i}=0$, respectively. Denote the distribution
functions of~$(B_{i},E_{i},S_{i})$ for $T_{i}=1$ and $T_{i}=0$ as
$f_{1}$ and $f_{0}$, respectively.

\subsection{Standard mixture joint model}
We first consider joint modeling of binding, expression and sequence
data without incorporating gene networks. We have the following
standard mixture joint model (SMJM):
%
\begin{equation}\label{SMJM}
f(B_{i},E_{i},S_{i}) = (1-\pi_{1})f_{0}(B_{i},E_{i},S_{i}) +\pi
_{1}f_{1}(B_{i},E_{i},S_{i}),
\end{equation}
where $\pi_{1}= \operatorname{Pr}(T_{i}=1)$ is the prior probability of gene $i$ being
a target. Note that it is the same for all the genes. We further
specify the conditional distribution $f_{j}=\phi(\cdot;\mu_{j},\Sigma
_{j})$, a multivariate normal density function with mean vector $\mu
_{j}$ and covariance matrix $\Sigma_{j}$ for $j=0,1$. Here we allow the
conditional covariance matrix $\Sigma_{j}$ to have a general structure,
that is, the three data sources can be correlated given $T_{i}$. A
special case is diagonal covariance matrix $\Sigma_{j}=\operatorname{Diag}(\sigma
_{B}^2,\sigma_{E}^2,\sigma_{S}^2)$, that is, the three data sources are
conditionally independent, as assumed in \citet{PanWeiKho08}. When
only one type of data, for example, gene expression data, is
considered, the conditional distributions $f_{0}$ and $f_{1}$ become
univariate normal density functions, and we call the corresponding
model ``standard mixture model''~(SMM).\vspace*{-2pt}

\begin{table}
\tabcolsep=4.5pt
\caption{Summary statistics of the two gene networks used
in the analysis}\label{tabnet}
\begin{tabular*}{\tablewidth}{@{\extracolsep{\fill}} l c c c c c c c@{}}
\hline
&&& \multicolumn{5}{c@{}}{\textbf{Percentiles of \# of direct
neighbors}}\\[-4pt]
&&& \multicolumn{5}{c@{}}{\hrulefill}\\
\textbf{Network} & \textbf{\# of nodes} &
\textbf{\# of edges} & \textbf{0\%} & $\bolds{25\%}$
& $\bolds{50\%}$ & $\bolds{75\%}$ & $\bolds{100\%}$ \\
\hline
Co-expression & 3,208 & \hphantom{0}86,791 & 1 & \hphantom{0}5
& \hphantom{0}20 & \hphantom{0}64 & 424 \\
Functional coupling (GO) & 1,644 & 116,422 & 1 & 48 & 102 & 249 & 708
\\
\hline
\end{tabular*}\vspace*{-3pt}
\end{table}

\subsection{MRF-based mixture joint model}
\label{chapter4sectionDMRF}
$\!\!$Because neighboring genes on a~network, for example, a co-expression or
functional coupling network, tend to be co-regulated by a TF and there
is more than one gene network available, each containing complementary
yet partial information about gene--gene interactions, it is desired to
incorporate multiple gene networks into joint modeling of genomic data.
Here we propose an MRF-based Mixture Joint Model (MRF-MJM) to
accomplish this goal. In contrast to assuming a priori i.i.d. gene
state $T_{i}$'s as in the SMJM, we model the state vector $\mathbf{T}=(T_{1},\ldots,T_{G})^{\prime}$ as MRFs defined on multiple neighborhood
systems, each corresponding to a gene network. Specifically, we propose
the following auto-logistic model for the distribution of $T_{i}$,
conditional on $T_{(-i)}=\{T_{l};l\neq i\}$:
%
\begin{eqnarray}\label{eqMRF}
\operatorname{logit} \operatorname{Pr}\bigl(T_{i}=1|T_{(-i)},\Phi\bigr)
& =& \operatorname{logit}
\operatorname{Pr}\bigl(T_{i}=1|T_{(\bigcup_{k=1}^{K}\partial i^{(k)})},\Phi\bigr)
\nonumber\\[-9pt]\\[-9pt]
&= & \gamma+ \sum_{k=1}^{K}\beta_{k}
\bigl[n_{i}^{(k)}(1)-n_{i}^{(k)}(0)\bigr]/m_{i}^{(k)},
\nonumber
\end{eqnarray}
where $\Phi=(\gamma,\beta_{1},\ldots,\beta_{K})$, $\gamma\in\mathbb
{R}$, $\beta_{k}\ge0$, $\partial i^{(k)}$ is the set of indices for
gene~$i$'s direct neighbors on network $\mathcal{G}_{k}$ for $k=1,\ldots
,K$, $n_{i}^{(k)}(j)$ is the\vadjust{\goodbreak} number of ge\-ne~$i$'s neighbors having
state $j$ on network $\mathcal{G}_k$ for $j=0,1$, and thus
$n_{i}^{(k)}(1)-\allowbreak n_{i}^{(k)}(0)=\sum_{l\in\partial i^{(k)}}(2T_{l}-1)$;
$m_{i}^{(k)}=n_{i}^{(k)}(0)+n_{i}^{(k)}(1)$ is the corresponding total
number of neighbors. The conditional probability of gene $i$ being a
target only depends on the states of its neighbors, as defined on the
$K$ networks, which is often referred to as the ``local dependency''
property. Note that we assume the contribution of each network to $
\operatorname{logit} \operatorname{Pr}(T_{i}=1|T_{(-i)},\Phi)$ is additive, weighted by the
nonnegative parameters $\beta_{k}$'s. Larger $\beta_{k}$ would induce
more similar states (target or nontarget) among neighboring genes on
network $\mathcal{G}_{k}$. In addition, the conditional distribution of
the observed data~$(B_{i},E_{i},S_{i})$ given $T_{i}$ is the same as
that in the SMJM.

The advantage of our proposed model is to combine all available gene
network information, and thus to boost the statistical power for
detecting target genes as much as possible. For example, as shown is
Figure~\ref{jointF1}, oraA is a true target that is not connected to
any other target genes in the GO-induced network, but is connected to
other targets in the co-expression network. As a result, in contrast to
using the GO-induced network alone, oraA's prior probability of being a
target can still be boosted by using the proposed model here to combine
both networks. Moreover, because $
[n_{i}^{(k)}(1)-\allowbreak n_{i}^{(k)}(0)]/m_{i}^{(k)}$ is always between
$-1$ and $1$, $\beta_{k}$'s are comparable and may be used to measure
how informative network $\mathcal{G}_{k}$ is. When $\beta_{1}=\cdots
=\beta_{K}=0$, the MRF-MJM is reduced to the SMJM. This can be seen by
noticing that $\operatorname{logit} \operatorname{Pr}(T_{i}=1|T_{(-i)},\Phi)=\gamma=
\operatorname{logit}\operatorname{Pr}(T_{i}=1)=\operatorname{logit}(\pi_{1})$, or, equivalently, $\pi
_{1}=\frac{e^{\gamma}}{1+e^{\gamma}}$, where $\pi_{1}$ is the prior
probability of being a target as defined in~(\ref{SMJM}) in the SMJM.

Singleton genes, that is, those without any neighbors in a network, are allowed
in the proposed MRF-MJM here. Denote $\mathcal{S}_{k}$ as the
set of indices for singletons in gene network $\mathcal{G}_{k}$. For
singleton gene $i\in\mathcal{S}_{k}$, we set $
[n_{i}^{(k)}(1)-n_{i}^{(k)}(0)]/ m_{i}^{(k)}=0$. If $i \in\bigcap
_{k=1}^{K} \mathcal{S}_{k}$, then $\operatorname{logit} \operatorname{Pr}(T_{i}=1|T_{(-i)},\Phi
)=\operatorname{logit}\operatorname{Pr}(T_{i}=1)=\gamma$.

Due to the unknown normalizing constant $C(\Phi)$ in the joint
distribution of $\mathbf{T}=(T_{1},\ldots,T_{G})'$, the likelihood
$l(\mathbf{T};\Phi)$ does not have a closed form. Instead, we propose
to use the \textit{pseudolikelihood} of \citet{Bes86}:
%
\begin{eqnarray}\label{eqchapter4pseudo}
pl(\mathbf{T};\Phi) &=& \prod_{i=1}^{G}p\bigl(T_{i}|T_{(\bigcup_{k=1}^{K}\partial
i^{(k)})},\Phi\bigr)\nonumber\\[-8pt]\\[-8pt]
&=&\prod_{i=1}^{G}\frac{\exp\{T_{i}(\gamma+ \sum
_{k=1}^{K}\beta_{k}[n_{i}^{(k)}(1)-n_{i}^{(k)}(0)
]/m_{i}^{(k)})\}}{1+\exp\{\gamma+ \sum_{k=1}^{K}\beta
_{k}[n_{i}^{(k)}(1)-n_{i}^{(k)}(0)]/m_{i}^{(k)}\}}.\nonumber
\end{eqnarray}
The maximizer of the pseudolikelihood was shown to be a consistent
estimator of the MRF parameters $\Phi$ [\citet{Win03}, page 272],
while \citet{RydTit98} showed that the pseudolikehood
$pl(\mathbf{T};\Phi)$ provides a good approximation to the genuine
likelihood $l(\mathbf{T};\Phi)$ in Bayesian hierarchical modeling as
adopted here. We found the approximation works well in our real data
analysis and simulation study.

Note that our proposed MRF defined on multiple neighborhoods is similar
to that used by \citet{DenCheSun04} in the context of protein
function prediction, rather than detection of the target genes of a TF here.

\subsection{Prior distributions}
\label{chapter4sectionprior} We use vague or noninformative prior
distributions. We denote by $\operatorname{MVN}(\bm{\mu},\Sigma)$ the
multivariate normal distribution with mean vector $\bm{\mu}$ and
covariance matrix $\Sigma$, and denote by $W((\rho R)^{-1},\rho)$ the
\textit{Wishart} distribution with mean vector $R^{-1}$. Reparameterize
the component-wise mean vector as $\bm{\mu
}_{1}=\bm{\mu}_{0}+\bm{\theta}$. We use the following priors for the
parameters in the conditional distribution of the observed data: $\bm
{\mu}_{0}\sim \operatorname{MVN}(\bm{0},\bmm{C})$, $\bm{\theta}\sim
\operatorname{MVN}(\bm{0},\bmm{C})I(\bm{\theta}>\bm{0})$, where
$\bmm{C}=\operatorname{diag}(10^{6},10^{6},10^{6})$; $\Sigma_{j}^{-1}
\sim W((3R)^{-1},3)$ for $j=0,1$, where $R$ is taken as the estimated
marginal covariance matrix of the three data sources whose off-diagonal
elements are close to zero. Since\vspace*{1pt} we have
$E(\Sigma_{j}^{-1})=R^{-1}$, $R$ is approximately the expected prior
variance of~$\Sigma_{j}$. This is considered as a very vague prior with
respect to the correlation parameters [\citet{CarLou09},
page~338]. For the SMJM, we have $\pi_{1}\sim
\operatorname{Beta}(1,1)$. For the MRF-MJM, we have $\gamma\propto1$
and $\beta_{k}\propto I(0 \le\beta_{k} < 6)$, $k=1,\ldots,K$.

\subsection{Statistical inference}
We carry out statistical inference in a fully Bayesian framework via
MCMC sampling. The MCMC algorithm for the SMJM can be implemented in
WinBUGS V1.40 [\citet{Spietal03}], while we wrote an R program to
implement the MCMC algorithm for the MRF-MJM. The WinBUGS code for the
SMJM is provided in the supplemental article [\citet{WeiPan}]. The
MCMC algorithm for the MRF-MJM can be found in the \hyperref[app]{Appendix}, and the R
program is available upon request.

We run three parallel chains of our MCMC algorithms starting from
different values, each run for 10,000 iterations after discarding the
first 5,000 as burn-in samples. We use the three parallel chains to
monitor convergence and obtain more stable posterior estimates by
combining the three chains. We use trace plots and the $\widehat{R}$
statistic of \citet{Dav92} to monitor the mixing of the Markov
chains; see Section \ref{sec43} and Supplemental Figure 2. The
posterior mean of any parameter based on combining 10,000 MCMC samples
after 5,000 burn-ins from each of the three chains is used as its point
estimate. In particular, we rank genes based on the posterior
probability of being a target
$\widehat{p_{i}}=\widehat{\operatorname{Pr}}(T_{i}=1|\mbox{Data})$. False
Discovery Rate (FDR) can be estimated based on $\widehat{p_{i}}$ as
discussed by \citet{WeiPan10}, which is not pursued in this study.

\section{Application to LexA data}
\subsection{Conditional independence assumption}
We applied the SMJM to jointly model the ChIP-chip binding, gene
expression and DNA sequence data. Table \ref{jointt1} shows the point
and interval estimates for the parameters in the conditional
correlation matrices of the three data sources. For the nontarget
%
\begin{table}
\tabcolsep=0pt
\caption{Posterior estimates for component-wise
(conditional) correlation matrices of binding (B), expression (E) and
sequence (S) data in the SMJM. Numbers in the parentheses are 95\%
credible intervals}\label{jointt1}
{\fontsize{8.5pt}{10pt}\selectfont{
\begin{tabular*}{\tablewidth}{@{\extracolsep{\fill}}l c l l c c l l@{}}
\hline
\multicolumn{4}{@{}c}{\textbf{Nontarget component}} &
\multicolumn{4}{c@{}}{\textbf{Target component}} \\[-4pt]
\multicolumn{4}{@{}c}{\hrulefill} &
\multicolumn{4}{c@{}}{\hrulefill} \\
& \multicolumn{1}{c}{\textbf{B}} & \multicolumn{1}{c}{\textbf{E}}
& \multicolumn{1}{c}{\textbf{S}} & & \multicolumn{1}{c}{\textbf{B}}
& \multicolumn{1}{c}{\textbf{E}} & \multicolumn{1}{c@{}}{\textbf{S}} \\
\hline
B& 1 & 0.013 ($-$0.027, 0.047) & $-$0.013 ($-$0.053, 0.023)
& B & 1 & \textbf{0.119 (0.034, 0.184)} & \textbf{0.475 (0.427, 0.513)} \\
E& & 1 & \hphantom{$-$}0.010 ($-$0.029, 0.045) & E & & 1 & 0.077 ($-$0.016, 0.147) \\
S& & & \hphantom{$-$}1 & S & & & 1 \\
\hline
\end{tabular*}}}
\end{table}
component, the three sources of data appeared to be independent with
each other. Interestingly, for the target component, binding and
sequence data were highly correlated, in contrast to the other two
pairs: binding and expression data, sequence and expression data, which
turned out to be only slightly correlated and independent,
respectively. This is consistent with the recent finding that LexA's
binding affinity to its regulatory targets depends on the extent to
which the binding site matches the consensus sequence for LexA
[\citet{ButZfuBus09}]. In addition, our results suggest that LexA
is quite efficient in repressing its target genes' expression: weak
binding only decreases its repression effect slightly.

\subsection{Predictive performance}
We evaluated the different methods' predictive performance by comparing
the ranks given by each method for 26 LexA's known and putative targets
annotated in RegulonDB (v6.4), as shown in Table \ref{jointt2}. Note
that known target genes of LexA were those experimentally verified via
binding of purified proteins, which was considered as ``strong''
evidence by RegulonDB [\citet{Gametal08}], whereas
putative target genes were those supported only by some ``weak''
evidence, for example, gene expression analysis or computational
prediction based on similarity to consensus sequence. Thus, evaluations
based on known targets are much more reliable than those based on
putative ones. As a result, we first focused on LexA's known targets.

%
\begin{sidewaystable}
\textwidth=\textheight
\tablewidth=\textwidth
\caption{Ranks given by various methods based on
posterior probabilities for known (marked by *) and putative target
genes of LexA annotated in RegulonDB. ``SMM'': standard mixture model;
``S'': SMJM with diagonal covariance; ``S.mul'': SMJM with general
covariance; ``co-exp'': co-expression network; ``GO'': functional
coupling network induced by GO}\label{jointt2}
{\fontsize{8.6pt}{9.5pt}\selectfont{
\begin{tabular*}{\tablewidth}{@{\extracolsep{\fill}}l r d{4.0} d{4.0}
d{4.0} d{4.0} d{4.0} r d{4.0} d{4.0} d{4.0} d{4.0} @{}}
\hline
& \multicolumn{4}{c}{\textbf{Expression}}
& & & \multicolumn{5}{c}{$\mbox{\textbf{Binding}}\bolds{+}\mbox{\textbf{Expression}}\bolds{+}
\mbox{\textbf{Sequence}}$} \\[-4pt]
& \multicolumn{4}{c}{\hrulefill} & & & \multicolumn{5}{c@{}}{\hrulefill}
\\
& & \multicolumn{3}{c}{\textbf{MRF-MJM}}
& \multicolumn{1}{c}{\multirow{2}{40pt}[-5pt]{\centering{\textbf{Binding} \textbf{SMM}}}}
& \multicolumn{1}{c}{\multirow{2}{40pt}[-5pt]{\centering{\textbf{Sequence} \textbf{SMM}}}}
& & & \multicolumn{3}{c@{}}{\textbf{MRF-MJM}}\\[-4pt]
& & \multicolumn{3}{c}{\hrulefill} & & & & &
\multicolumn{3}{c@{}}{\hrulefill}\\
\textbf{Targets} & \multicolumn{1}{c}{\textbf{SMM}}
& \multicolumn{1}{c}{\textbf{co-exp}} & \multicolumn{1}{c}{\textbf{GO}}
& \multicolumn{1}{c}{\textbf{co-exp${}\bolds{+}{}$GO}} &
\multicolumn{1}{c}{} & \multicolumn{1}{c}{}
& \multicolumn{1}{c}{\textbf{S}} & \multicolumn{1}{c}{\textbf{S.mul}}
& \multicolumn{1}{c}{\textbf{co-exp}} & \multicolumn{1}{c}{\textbf{GO}}
& \multicolumn{1}{c@{}}{\textbf{co-exp${}\bolds{+}{}$GO}}\\
\hline
umuD* & 1 & 1 & 1 & 1 & 1 & 1 & 1 & 1 & 1 & 1 & 1 \\
recN* & 1 & 1 & 1 & 1 & 1 & 1 & 1 & 1 & 1 & 1 & 1 \\
recA* & 1 & 1 & 1 & 1 & 1 & 1 & 1 & 1 & 1 & 1 & 1 \\
lexA* & 1 & 1 & 1 & 1 & 1 & 1 & 1 & 1 & 1 & 1 & 1 \\
dinI* & 1 & 1 & 1 & 1 & 1 & 48 & 1 & 1 & 1 & 1 & 1 \\
ydjM* & 1 & 1 & 1 & 1 & 1 & 70 & 1 & 1 & 1 & 1 & 1 \\
oraA* & 1 & 1 & 1 & 1 & 82 & 1\mbox{,}206 & 1 & 1 & 1 & 1 & 1 \\
polB* & 1 & 1 & 1 & 1 & 156 & 153 & 1 & 1 & 1 & 1 & 1 \\
umuC* & 1 & 1 & 1 & 1 & 192 & 3\mbox{,}500 & 1 & 1 & 1 & 1 & 1 \\
sulA & 1 & 1 & 1 & 1 & 1 & 1 & 1 & 1 & 1 & 1 & 1 \\
ssb & 129 & 1 & 133 & 1 & 1 & 1 & 1 & 1 & 1 & 1 & 1 \\
ruvA* & 146 & 1 & 133 & 1 & 127 & 1 & 1 & 1 & 1 & 1 & 1 \\
uvrA* & 163 & 134 & 159 & 133 & 1 & 1 & 1 & 1 & 1 & 1 & 1 \\
uvrB* & 172 & 134 & 175 & 133 & 1 & 1 & 1 & 1 & 1 & 1 & 1 \\
t150 & 172 & 176 & 167 & 169 & 2\mbox{,}118 & 50 & 173 & 215 & 178 & 172 & 174 \\
dinF* & 216 & 182 & 214 & 178 & 2\mbox{,}471 & 1 & 1 & 145 & 1 & 1 & 1 \\
uvrD* & 245 & 259 & 249 & 261 & 262 & 1 & 1 & 1 & 1 & 1 & 1 \\
ruvB* & 311 & 226 & 313 & 231 & 2\mbox{,}118 & 1\mbox{,}456 & 644 & 576 & 367 & 614 & 373 \\
dinG* & 450 & 311 & 439 & 314 & 96 & 136 & 168 & 168 & 142 & 166 & 144 \\
rpsU & 1\mbox{,}190 & 1\mbox{,}810 & 2\mbox{,}694 & 2\mbox{,}445 & 470 & 1\mbox{,}091 & 886 & 955 & 1\mbox{,}021 & 1\mbox{,}105 & 1\mbox{,}266 \\
phrB & 1\mbox{,}738 & 2\mbox{,}858 & 2\mbox{,}819 & 3\mbox{,}137 & 1\mbox{,}334 & 531 & 1\mbox{,}460 & 1\mbox{,}686 & 2\mbox{,}031 & 1\mbox{,}898 & 2\mbox{,}154 \\
uvrC & 2\mbox{,}534 & 1\mbox{,}401 & 2\mbox{,}715 & 1\mbox{,}467 & 3\mbox{,}022 & 3\mbox{,}334 & 3\mbox{,}080 & 2\mbox{,}980 & 1\mbox{,}937 & 2\mbox{,}978 & 1\mbox{,}956 \\
dnaG & 3\mbox{,}060 & 3\mbox{,}119 & 3\mbox{,}100 & 3\mbox{,}266 & 2\mbox{,}471 & 781 & 2\mbox{,}831 & 3\mbox{,}169 & 2\mbox{,}897 & 2\mbox{,}978 & 3\mbox{,}087 \\
rpoD & 3\mbox{,}336 & 3\mbox{,}727 & 2\mbox{,}422 & 2\mbox{,}969 & 2\mbox{,}471 & 791 & 2\mbox{,}622 & 3\mbox{,}169 & 2\mbox{,}897 & 2\mbox{,}199 & 2\mbox{,}685 \\
ftsK* & 3\mbox{,}723 & 3\mbox{,}583 & 2\mbox{,}313 & 2\mbox{,}727 & 75 & 128 & 169 & 171 & 180 & 166 & 174 \\
uvrY & 3\mbox{,}723 & 3\mbox{,}472 & 2\mbox{,}313 & 2\mbox{,}727 & 3\mbox{,}022 & 3\mbox{,}500 & 3\mbox{,}080 & 2\mbox{,}964 & 3\mbox{,}173 & 2\mbox{,}789 & 2\mbox{,}884 \\
[4pt]
\# Tied rank 1 & 128 & 133 & 132 & 132 & 53 & 36 & 145 & 144 & 141 & 145 & 143 \\
\hline
\end{tabular*}}}\vspace*{-4pt}
\end{sidewaystable}

In general, incorporating gene networks and combining additional types
of genomic data increased the chance of detecting the true targets as
compared to using a single type of genomic data alone; this was
evidenced by higher, in some cases substantially higher, ranks based on
the integrative analyses\vadjust{\goodbreak} than those based on using binding, expression,
or sequence data alone. When network information was not utilized, many
of LexA's known targets did not have consistently high ranking based on
any of the three genomic data sources alone. For example, oraA and dinF
were ranked 82nd and 2,471st, respectively, based on binding data alone,
while they were ranked 1,206th and tied first, respectively, based on
sequence data alone. In contrast, the majority of LexA's known targets
(14 out of 17) were boosted to a highest rank, that is, tied at the
first with posterior probability equal to 1, by combining all three
sources of genomic data. On the other hand, incorporating multiple gene
networks into modeling of a single source of genomic data also led to
dramatic rank improvement. For example, ruvA and uvrB were ranked 146th
and 172nd based on expression data alone, but with the incorporation of
gene networks their ranks improved to a tied first and 133rd,
respectively. This was achieved without the aid of additional genomic
data such as binding and sequence data, demonstrating the extra power
gained by incorporating multiple gene networks. Compared with the
significant rank improvement by the network-based analyses of a single
type of genomic data, integrating multiple networks with all three
sources of genomic data resulted in less dramatic improvement in
predictive performance over joint modeling of genomic data only,
possibly because the latter already had very high predictive power.

In addition, several features are noticeable. First, using a general
conditional covariance structure in the SMJM did not lead to improved
rankings as compared to using a diagonal conditional covariance
structure. As a result, we used a diagonal conditional covariance
structure in all MRF-based analyses for better predictive performance.
Second, when integrating more than one gene network, we observed that
the predictive performance tended to be compromised, that is, the ranks
based on both networks were often between those based on the
co-expression network alone and those based on the GO-induced network
alone. For example, dinG was ranked 142nd and 166th by the
co-expression network-based and GO network-based MRF-MJM, respectively,
whereas it was ranked 144th by the MRF-MJM that incorporated both
networks. Third, as shown in Table \ref{jointt3}, the relative
magnitude of the weights $\beta$'s for the two gene networks in the
MRF-MJM were quite consistent: the co-expression network had higher
weight than did the GO-induced network. Given the observation that the
co-expression network-based analyses tended to lead to higher ranks
than the GO network-based analyses, especially for modeling the gene
expression alone, $\beta$ may be used to measure how ``good'' a gene
network is. One possible reason why the GO-induced gene network was not
as good as the co-expression network was that the former network was
much more densely connected, as illustrated by Table \ref{tabnet} and
Figure \ref{jointF1}, resulting in higher probability of target and
nontarget genes being direct neighbors in the network, and thus,
reduced power of the network-based methods.

\begin{table}
\caption{Posterior means of parameters in the MRF-MJM
(B: Binding; E: Expression; S: Sequence)}\label{jointt3}
\begin{tabular*}{\tablewidth}{@{\extracolsep{\fill}}l c d{2.2} c c @{}}
\hline
\textbf{Genomic data} & \textbf{Networks}
& \multicolumn{1}{c}{$\bolds{\gamma}$}
& \multicolumn{1}{c}{$\bolds{\beta_{\mathrm{co}\mbox{-}\mathrm{expression}}}$}
& \multicolumn{1}{c@{}}{$\bolds{\beta_{\mathrm{GO}}}$} \\
\hline
$\mbox{B}+\mbox{E}+\mbox{S}$ & Co-expression & -1.33 & 1.16 & -- \\
& GO & -1.72 & -- & 0.84 \\
& Co-expression${} + {}$GO & -1.20 & 1.07 & 0.61 \\ [6pt]
E & Co-expression & -0.88 & 1.35 & -- \\
& GO & -1.30 & -- & 0.99 \\
& Co-expression${} + {}$GO & -0.73 & 1.26 & 0.71 \\
\hline
\end{tabular*}
\vspace*{-2pt}
\end{table}

Our joint modeling analyses also enabled us to potentially distinguish
true targets of LexA from false positives in the putative target gene
list. Among the nine putative targets, three genes---sulA, ssb and
t150---were consistently highly ranked by various models based on different
data sources, and thus were very likely to be true targets of LexA. In
contrast, the rest of the six putative targets had consistent low
rankings, suggesting that they were likely to be false positive target
genes. Interestingly, as shown in Figure~\ref{jointF1}, sulA and ssb
were both direct neighbors of some known targets of LexA in both
co-expression and GO gene networks, whereas none and only three of the
six genes that were likely to be false positives were direct neighbors
of known targets in the co-expression and GO network, respectively.

We noticed that there were quite a few genes with tied rank ones,
ranging from 36 to 145 genes across different data sources and networks
(Table \ref{jointt2}). Those genes' genomic data, that is, binding,
expression or sequence scores, were among the highest, and, as a result
of their falling in the farthest right tail of the mixture
distribution, the MCMC ended up with always drawing $T_{i}=1$ for those
genes across the entire finite iterations. It is noteworthy that the
number of tied ones mainly depended on how much the two mixture
components $f_{0}$ and $f_{1}$ in (\ref{SMJM}) were separated.
Specifically, the expression data, whose two components had the best
separation among the three data sources, led to 128 tied ones, whereas
the sequence data, least separated, had~36 tied ones. Combining the
three sources resulted in a higher number of tied ones than did any
single source alone. Ties at other ranks were possible due to finite
iterations of the MCMC.\vspace*{-2pt}

\subsection{Convergence diagnostics and sensitivity
analysis}\label{sec43}
Given the large number of parameters, we only visually check the MCMC
convergence for the mixture component and MRF parameters, that is, $\bm
{\mu}_{0}$, $\bm{\mu}_{1}$, $\Sigma$ and $\Phi$, whose convergence
should also indicate that of the latent state vector $\mathbf{T}=(T_{1},\ldots,T_{G})^{\prime}$. The trace plots did not reveal any
convergence problems and the\vadjust{\goodbreak} $\widehat{R}$ statistics of \citet{Dav92} were all close to 1, indicating that the multiple chains
mixed with each other and converged by 5,000 iterations; see
Supplemental Figure 2. The posterior probabilities $p(T_i=1)$ based on
each individual Markov chain showed very little difference;
nevertheless, we combined the MCMC samples from the three chains to
obtain more stable posterior estimates.

In our proposed network-based joint model, we used noninformative or
vague priors for the mixture component and MRF parameters as described
in Section~\ref{chapter4sectionprior}, whereas we used gene networks as
informative priors for the latent state vector $\mathbf{T}$. As
evidence of minimal influence of the adopted priors on the posterior
estimates of the mixture model parameters, the resulting posterior
means in the SMJM were very close to the maximum likelihood estimates
(MLEs) obtained via the EM algorithm [\citet{DemLaiRub77}]
(results not shown). On the other hand, we performed a sensitivity
analysis to investigate how robust the network-based results were to
potential incomplete/misspecified gene networks. Specifically, we
applied the two co-expression networks with correlation coefficient
cutoffs of 0.60 and 0.70 to the expression data alone as well as joint
modeling of the three data sources, and compared the results to those
based on the co-expression network with the cutoff of 0.65.
Supplemental Figure 1 shows the three subnetworks, consisting of LexA's
known and putative target genes, from the co-expression networks with
the cutoffs of 0.60, 0.65 and 0.70, respectively. The genes that formed
a connected subnetwork were the same for the cutoffs 0.60 and 0.65,
whereas ydjM and ssb became singletons in the subnetwork with the
cutoff of 0.70. As shown in Supplemental Table 1, in spite of
quantitative difference in the known target genes' ranks based on the
co-expression networks with different cutoffs, the network-based
analyses consistently improved the predictive performance compared with
the analyses of genomic data alone. As of the singleton genes ydjM and
ssb in the co-expression subnetwork with the cutoff of 0.70, only ssb
had slightly lower rank based on the network-based analysis of
expression data and all other network-based analyses resulted in tied
first for both genes due to strong genomic data signals. Our results
demonstrate that the network-based methods are reasonably robust to
misspecification of the network structures, consistent with previous
sensitivity analysis results [Wei and Pan (\citeyear{WeiPan08N1},
\citeyear{WeiPan10}), \citet{WeiLi08}].

\section{Simulation study}
To further evaluate the conditional independence assumption and the
effects of integrating multiple networks and diverse types of genomic
data, we conducted a simulation study that mimicked the real data: the
co-expression network was more informative than the GO-induced network
and the conditional covariance matrices in the simulation model were
based on those estimated from the real data. Specifically, the latent
states vector $\mathbf{T}$ was based on the fitted MRF-MJM that
incorporated both gene networks, while, given $\mathbf{T}$, the
observed genomic data were generated based on the fitted SMJM with a\vadjust{\goodbreak}
general conditional covariance structure. We let the top 487 genes,
which are $\widehat{\pi}_{1}=13\%$ of the total 3,779 genes, in the
fitted MRF-MJM that incorporated both networks be targets ($T_{i}=1$)
and the rest of the 3,292 genes be nontargets ($T_{i}=0$). Note that the
posterior means for the weight parameters
$\beta_{\mathrm{co}\mbox{-}\mathrm{exp}}$ and $\beta _{\mathrm{GO}}$ were 1.06
and 0.61, respectively. Given $\mathbf{T}$, we simulated the binding,
expression and sequence data from the fitted conditional normal
distributions with nontarget mean vector $\widehat{\bm{\mu
}}_{0}=(0.11,0.02,13.35)'$, target mean vector $\widehat{\bm{\mu
}}_{1}=(0.50,0.26,14.58)'$ and covariance matrices corresponding to the
correlation matrices in Table~\ref{jointt1}.

\begin{figure}
\begin{tabular}{@{}c@{\hspace*{6pt}}c@{}}

\includegraphics{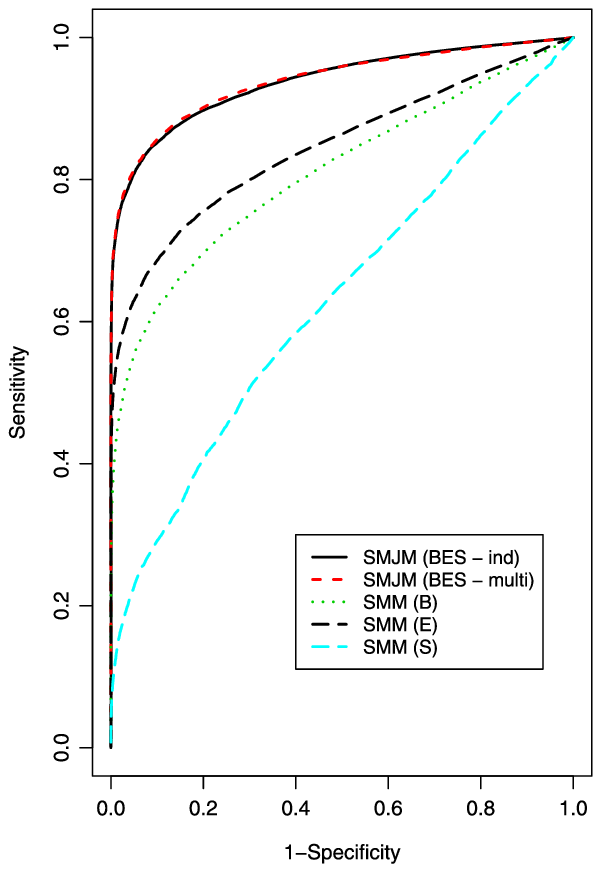}
 & \includegraphics{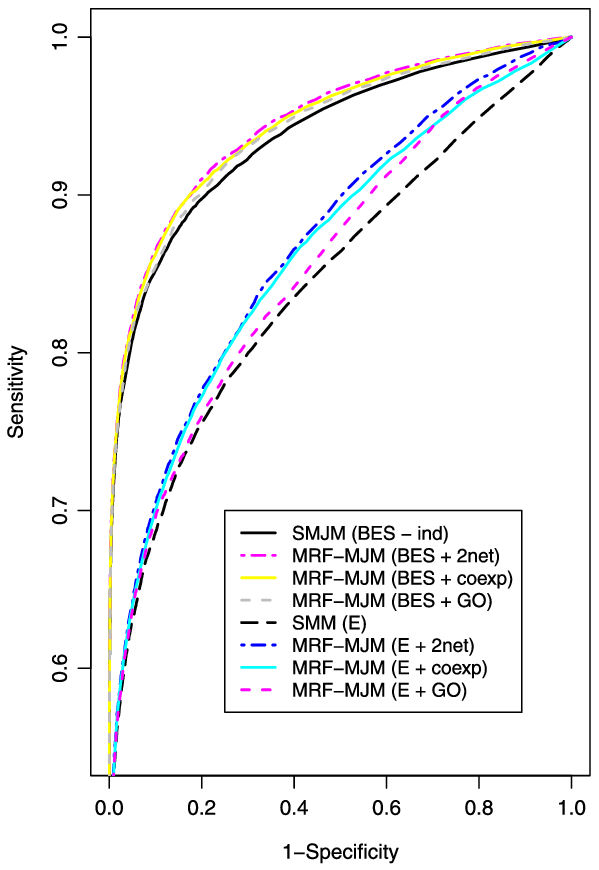} \\
(a) & (b)
\end{tabular}
\caption{ROC curves (averaged over 20 simulated data sets) for
\textup{(a)}
modeling genomic data alone (``B'' for binding, ``E'' for expression,
``S'' for sequence, ``multi'' and ``ind'' for a~general and a diagonal
conditional covariance structure, resp.) and \textup{(b)} MRF-MJM (``GO''
for GO-induced network, ``coexp'' for co-expression network, ``2net''
for both networks).}
\label{simuind}
\end{figure}

We simulated 20 data sets and applied the SMJM with an either general
or diagonal conditional covariance structure and the MRF-MJM to each of
the data sets. We used the ROC curves to compare the predictive
performance. Figure \ref{simuind} shows the ROC curves averaged across
the 20 simulated data sets. When no network information was utilized,
as shown in Figure~\ref{simuind}(a), joint modeling of the three data
sources, that is, the SMJM with either covariance structure, had much
higher predictive power than using a~single source of genomic data.\vadjust{\goodbreak} On
the other hand, although the simulated binding and sequence data were
considerably correlated for the target genes, assuming conditional
dependence by adopting a general covariance structure hardly made any
difference in terms of predictive power. This may be explained by the
fact that the sequence data were the least informative among the three
data sources, as suggested by the ROC curves, making the strong
correlation between the binding and sequence data among the target
genes much less important in terms of predictive power.

Incorporating gene networks via the MRF-MJM led to dominating ROC
curves over those based on genomic data alone, as shown in Figure \ref
{simuind}(b). Consistent with the real data analysis results, the
improved power by the MRF-MJM was more dramatic for using the
expression data alone than joint modeling of the three data sources. As
pointed out by \citet{WeiPan10}, the posterior probability of being a
target in the MRF-based mixture models was jointly determined by the
prior probability and the likelihood function, which depended on the
gene networks and the observed genomic data, respectively. When the
likelihood was very informative, such as the one for joint modeling of
the three data sources here, it might dominate the prior probabilities,
making the contribution of the gene networks less significant. In
addition, when only one network was incorporated, the ROC curve for the
co-expression network dominated that for the GO network, which was true
in both scenarios, using expression data alone or combining three data
sources, suggesting that the weight parameter $\beta$ can be useful in
comparing the ``informativeness'' of different gene networks. Finally,
incorporating both networks resulted in improved predictive performance
over using a~single network, especially the GO network, demonstrating
the flexibility and efficiency gains with the proposed MRF-MJM for
integrating multiple gene networks.\looseness=-1

\section{Discussion}
We have presented a flexible and powerful mixture model, based on the
use of multiple MRFs, for integrating diverse types of genomic data and
multiple gene networks to identify regulatory target genes of a TF.
Rather than assuming conditional independence of ChIP-chip binding,
gene expression and DNA sequence data, we allow multiple sources of
data to be conditionally correlated. Due to a fully Bayesian approach,
inference about model parameters can be easily carried out based on
MCMC samples. Application to the LexA data, together with simulation
studies, demonstrates the utility and statistical efficiency gains with
the proposed joint model. An interesting biological finding is that the
binding and sequence data were highly correlated for target genes only,
which helps elucidate the regulation mechanism of LexA, an important TF
involved in DNA repair in \mbox{E. coli}. Interestingly, ignoring the
conditional correlations even led to slightly improved predictive
performance. Our simulation study that mimicked the LexA real data
confirmed that incorrectly assuming conditional independence did not
result in deteriorated performance, possibly due to simpler models as
well as only moderate predictive power of the sequence data. Further
study on this problem is needed.

Although our application concerns identification of target genes of a
TF in \mbox{E. coli}, it may be possible to adapt the proposed method to
address other problems for other organisms, for example, identifying
genes predisposed to complex human diseases by integrating multiple
types of data such as SNP, epigenomic, gene expression, proteomic,
metabolomic data and gene networks/pathways. It has been recently
proposed to incorporate a single gene network into analysis of
genome-wide association study (GWAS) data via a MRF model
[\citet{CheChoZha11}]. In light of our
study here, it would be interesting to consider multiple gene networks
in network-based analysis of GWAS.

Based on the LexA data, we found that combining both gene networks
might result in compromised predictive performance. This raises a
question: shall we integrate as many gene networks as possible or
choose to use the ``best'' gene network? If the former, as demonstrated
by the simulation results, the MRF-MJM provides a very flexible and
efficient framework to combine multiple networks by down-weighting more
noisy ones. If the latter, how to compare gene networks is still an
open question. A possible perspective is to look at the structural or
topological differences between the networks. For example, as
illustrated by Table \ref{tabnet} and Figure \ref{jointF1}, the
GO-induced network may be too dense, directly connecting many target
and nontarget genes, and thus is less preferred compared to the
co-expression network. On the other hand, the weight parameter $\beta$
in the MRF-MJM has been demonstrated, by analyses of the LexA data as
well as the simulation results, to be a promising criterion for
quantitative comparison of gene networks. Nevertheless, considering
that each of the gene networks contains partial yet complementary
information about gene--gene interactions, integrating multiple networks
is likely to achieve higher predictive power on average, for example,
as measured by the area under the ROC curve (AUC). This could be a
direction of future research.

While discrete MRFs were employed here to incorporate multiple gene
networks, Gaussian MRFs [Wei and Pan (\citeyear{WeiPan08N1}, \citeyear{WeiPan10})] could be similarly
used. However, unlike $[n_{i}^{(k)}(1)-n_{i}^{(k)}(0)
]/m_{i}^{(k)}$ in (\ref{eqMRF}), which is always between $-1$ and $1$,
the range of a similar term based on the Gaussian MRF would be the real
line. As a result, it is unclear how to effectively assign weights to
different networks based on the use of multiple Gaussian MRFs. This,
together with assigning weights to different genomic data sources,
would be an interesting topic for future investigation.\vspace*{-2pt}

\begin{appendix}\label{app}
\section*{Appendix}\vspace*{-2pt}
\subsection{MCMC algorithm for the MRF-MJM}
We denote by $(\alpha|\ldots)$ the full conditional of $\alpha$, that
is, the distribution of $\alpha$ conditional on everything else\vadjust{\goodbreak} in the
model. In addition, we denote by $\operatorname{MVN}(\bm{\mu},\bm{\Sigma})$ the
multivariate normal distribution with mean vector $\bm{\mu}$ and
covariance matrix $\bm{\Sigma}$,\vspace*{1pt} by $\phi(\cdot;\bm{\mu},\bm{\Sigma})$ the
corresponding density function, and by $W((\rho R)^{-1},\rho)$ the
Wishart distribution with mean $R^{-1}$. The observed data are denoted
as $\mathbf{x}=\{x_{i}=(B_{i},E_{i},S_{i})^{\prime}; i=1,\ldots,G
\}$. Model specification and prior distributions for the MRF-MJM can be
found in Sections \ref{chapter4sectionDMRF} and \ref
{chapter4sectionprior}. In particular,~$p(\mathbf{T}|\Phi)$ is
specified by the pseudolikelihood (\ref{eqchapter4pseudo}). As
detailed below, we use Metropolis with Gibbs sampling to update $\Phi$.
The anxiliary variable-based Metropolis--Hastings algorithm of
\citet{Mlletal06} could be used to update $\Phi$ in the presence of
the unknown normalizing constant $C(\Phi)$, which could, however,
substantially slow down the computation, and is not pursued here.

The joint posterior distribution is
\begin{eqnarray*}
&&(\mathbf{T},\bm{\mu}_{0},\bm{\theta},\bm{\Sigma}_0,\bm{\Sigma}_{1},\Phi
|\mathbf{x}) \\
&&\qquad\propto p(\mathbf{x}|\mathbf{T},\bm{\mu}_{0},\bm{\theta
},\bm{\Sigma}_0,\bm{\Sigma}_{1})
p(\mathbf{T}|\Phi)p(\bm{\mu}_{0})p(\bm
{\theta})p(\bm{\Sigma}_{0})p(\bm{\Sigma}_{1})p(\Phi)\mbox{:}
\end{eqnarray*}
\begin{itemize}
\item Update $\bm{\mu}_{0}$ by Gibbs sampling with the proposal given by
\[
(\bm{\mu}_{0}|\ldots) \sim
\operatorname{MVN}\biggl((n_{0}\bm{\Sigma}_{0}^{-1}+\bmm{C}^{-1})^{-1}\bm{\Sigma}_{0}^{-1}\sum_{\{i\dvtx T_{i}=0\}
}x_{i},(n_{0}\bm{\Sigma}_{0}^{-1}+\bmm{C}^{-1})^{-1}\biggr),
\]
where $n_{0} = |\{ i\dvtx T_{i}=0\}|$.

\item Update $\bm{\theta}$ by Gibbs sampling with the proposal given by
\begin{eqnarray*}
(\bm{\theta}|\ldots) &\sim& \operatorname{MVN}\biggl(\!(n_{1}\bm{\Sigma}_{1}^{-1}\,{+}\,
    \bmm{C}^{-1})^{-1}\bm{\Sigma}_{1}^{-1}\!\!\sum_{\{i\dvtx T_{i}=1\}}\!\!(x_{i}\,{-}\,
    \bm{\mu}_{0}),(n_{1}\bm{\Sigma}_{1}^{-1}\,{+}\,\bmm{C}^{-1})^{-1}\!\biggr)\\
&&{}\times I(\bm{\theta}>0),
\end{eqnarray*}
where $n_{1} = |\{ i\dvtx T_{i}=1\}|$.

\item Update $\bm{\Sigma}_{j}$, for $j=0,1$, by Gibbs sampling with the
proposal given by
\[
(\bm{\Sigma}_{j}^{-1}|\ldots) \sim W\biggl(\biggl(\sum_{\{i\dvtx T_{i}=j\}
}(x_{i}-\bm{\mu}_{j})(x_{i}-\bm{\mu}_{j})^{\prime}+3R\biggr)^{-1},n_{j}+3\biggr),
\]
where $\bm{\mu}_1=\bm{\mu}_0+\bm{\theta}$.

\item Update $T_{i}$ by Gibbs sampling with proposal given by
\[
(T_{i}|\ldots) \sim\operatorname{Bernoulli}\biggl(\frac{d}{1+d}\biggr),
\]
where $d=\exp\{\gamma+ \sum_{k=1}^{K}\beta_{k}
[n_{i}^{(k)}(1)-n_{i}^{(k)}(0)]/m_{i}^{(k)}\}\frac{\phi
(x_{i};\bm{\mu}_{1},\bm{\Sigma}_{1})}{\phi(x_{i};\bm{\mu}_{0},\bm{\Sigma
}_{0})}$.\vspace*{2pt}

\item Update $\Phi=(\gamma,\beta_{1},\ldots,\beta_{K})$ using a random
walk Metropolis algorithm with Gaussian proposal, which has diagonal
covariance matrix. The acceptance ratio\vadjust{\goodbreak} is calculated using the full
conditional of $\Phi$, which is proportional to
\[
\frac{\exp\{ n_{1}\gamma+ \sum_{j=0}^{1}\sum_{i\dvtx T_{i}=j}\sum
_{k=1}^{K}\beta_{k}n_{i}^{(k)}(j)/m_{i}^{(k)}\}}{\prod
_{i=1}^{G}\{\exp(\sum_{k=1}^{K}\beta
_{k}n_{i}^{(k)}(0)/m_{i}^{(k)}) + \exp(\gamma+ \sum_{k=1}^{K}\beta
_{k}n_{i}^{(k)}(1)/m_{i}^{(k)}) \}}.
\]
The Gaussian proposal was tuned such that the acceptance rate was
around 0.23, the optimal one [\citet{CarLou09}, page 131].
\end{itemize}
\end{appendix}

\section*{Acknowledgments}

The authors are grateful to two anonymous reviewers and the Editor and
for their helpful and constructive comments that improved the
presentation of the paper.

\begin{supplement}[id=suppA]
\stitle{Supplemental tables and figures}
\slink[doi]{10.1214/11-AOAS502SUPP} 
\slink[url]{http://lib.stat.cmu.edu/aoas/502/supplement.pdf}
\sdatatype{.pdf}
\sdescription{WinBUGS codes, results for sensitivity analysis and MCMC
convergence diagnostics plots can be found in the supplemental article.}
\end{supplement}


\printaddresses


\begin{thebibliography}{40}

\bibitem[\protect\citeauthoryear{Ashburner et~al.}{2000}]{Ashetal00}
\begin{barticle}[pbm]
\bauthor{\bsnm{Ashburner},~\bfnm{M.}\binits{M.}},
  \bauthor{\bsnm{Ball},~\bfnm{C.~A.}\binits{C.~A.}},
  \bauthor{\bsnm{Blake},~\bfnm{J.~A.}\binits{J.~A.}},
  \bauthor{\bsnm{Botstein},~\bfnm{D.}\binits{D.}},
  \bauthor{\bsnm{Butler},~\bfnm{H.}\binits{H.}},
  \bauthor{\bsnm{Cherry},~\bfnm{J.~M.}\binits{J.~M.}},
  \bauthor{\bsnm{Davis},~\bfnm{A.~P.}\binits{A.~P.}},
  \bauthor{\bsnm{Dolinski},~\bfnm{K.}\binits{K.}},
  \bauthor{\bsnm{Dwight},~\bfnm{S.~S.}\binits{S.~S.}},
  \bauthor{\bsnm{Eppig},~\bfnm{J.~T.}\binits{J.~T.}},
  \bauthor{\bsnm{Harris},~\bfnm{M.~A.}\binits{M.~A.}},
  \bauthor{\bsnm{Hill},~\bfnm{D.~P.}\binits{D.~P.}},
  \bauthor{\bsnm{Issel-Tarver},~\bfnm{L.}\binits{L.}},
  \bauthor{\bsnm{Kasarskis},~\bfnm{A.}\binits{A.}},
  \bauthor{\bsnm{Lewis},~\bfnm{S.}\binits{S.}},
  \bauthor{\bsnm{Matese},~\bfnm{J.~C.}\binits{J.~C.}},
  \bauthor{\bsnm{Richardson},~\bfnm{J.~E.}\binits{J.~E.}},
  \bauthor{\bsnm{Ringwald},~\bfnm{M.}\binits{M.}},
  \bauthor{\bsnm{Rubin},~\bfnm{G.~M.}\binits{G.~M.}} \AND
  \bauthor{\bsnm{Sherlock},~\bfnm{G.}\binits{G.}}
(\byear{2000}).
\btitle{Gene ontology: Tool for the unification of biology. The Gene Ontology
  Consortium}.
\bjournal{Nat. Genet.}
\bvolume{25}
\bpages{25--29}.
\bid{doi={10.1038/75556}, issn={1061-4036}, mid={NIHMS269796}, pmcid={3037419},
  pmid={10802651}}
\bptok{imsref}%
\end{barticle}
\endbibitem

\bibitem[\protect\citeauthoryear{Bailey and Elkan}{1995}]{BaiElk95}
\begin{barticle}[auto:STB|2011/09/12|07:03:23]
\bauthor{\bsnm{Bailey},~\bfnm{T.~L.}\binits{T.~L.}} \AND
  \bauthor{\bsnm{Elkan},~\bfnm{C.}\binits{C.}}
(\byear{1995}).
\btitle{Unsupervised learning of multiple motifs in biopolymers using EM}.
\bjournal{Machine Learning}
\bvolume{21}
\bpages{51--80}.
\bptok{imsref}%
\end{barticle}
\endbibitem

\bibitem[\protect\citeauthoryear{Besag}{1986}]{Bes86}
\begin{barticle}[mr]
\bauthor{\bsnm{Besag},~\bfnm{Julian}\binits{J.}}
(\byear{1986}).
\btitle{On the statistical analysis of dirty pictures}.
\bjournal{J. Roy. Statist. Soc. Ser. B}
\bvolume{48}
\bpages{259--302}.
\bid{issn={0035-9246}, mr={0876840}}
\bptok{imsref}%
\end{barticle}
\endbibitem

\bibitem[\protect\citeauthoryear{Brown and Jurisica}{2005}]{BroJur05}
\begin{barticle}[pbm]
\bauthor{\bsnm{Brown},~\bfnm{Kevin~R.}\binits{K.~R.}} \AND
  \bauthor{\bsnm{Jurisica},~\bfnm{Igor}\binits{I.}}
(\byear{2005}).
\btitle{Online predicted human interaction database}.
\bjournal{Bioinformatics}
\bvolume{21}
\bpages{2076--2082}.
\bid{doi={10.1093/bioinformatics/bti273}, issn={1367-4803}, pii={bti273},
  pmid={15657099}}
\bptok{imsref}%
\end{barticle}
\endbibitem

\bibitem[\protect\citeauthoryear{Butala, Zfur-Bertok and
  Busby}{2009}]{ButZfuBus09}
\begin{barticle}[auto:STB|2011/09/12|07:03:23]
\bauthor{\bsnm{Butala},~\bfnm{M.}\binits{M.}},
  \bauthor{\bsnm{Zfur-Bertok},~\bfnm{D.}\binits{D.}} \AND
  \bauthor{\bsnm{Busby},~\bfnm{S.~J.~W.}\binits{S.~J.~W.}}
(\byear{2009}).
\btitle{The bacteria LexA transcriptional repressor}.
\bjournal{Cell. Mol. Life Sci.}
\bvolume{66}
\bpages{82--93}.
\bptok{imsref}%
\end{barticle}
\endbibitem

\bibitem[\protect\citeauthoryear{Carlin and Louis}{2009}]{CarLou09}
\begin{bbook}[mr]
\bauthor{\bsnm{Carlin},~\bfnm{Bradley~P.}\binits{B.~P.}} \AND
  \bauthor{\bsnm{Louis},~\bfnm{Thomas~A.}\binits{T.~A.}}
(\byear{2009}).
\btitle{Bayesian Methods for Data Analysis},
\bedition{3rd} ed.
\bpublisher{CRC Press}, \baddress{Boca Raton, FL}.
\bid{mr={2442364}}
\bptnote{check year}%
\bptok{imsref}%
\end{bbook}
\endbibitem

\bibitem[\protect\citeauthoryear{Chen, Cho and Zhao}{2011}]{CheChoZha11}
\begin{barticle}[auto:STB|2011/09/12|07:03:23]
\bauthor{\bsnm{Chen},~\bfnm{M.}\binits{M.}},
  \bauthor{\bsnm{Cho},~\bfnm{J.}\binits{J.}} \AND
  \bauthor{\bsnm{Zhao},~\bfnm{H.}\binits{H.}}
(\byear{2011}).
\btitle{Incorporating biological pathways via a Markov random field model in
  genome-wide association studies}.
\bjournal{PLoS Genet.}
\bvolume{7}
\bpages{e1001353}.
\bid{doi={10.1371/journal.pgen.1001353}}
\bptok{imsref}%
\end{barticle}
\endbibitem

\bibitem[\protect\citeauthoryear{Cirz et~al.}{2005}]{Ciretal05}
\begin{barticle}[pbm]
\bauthor{\bsnm{Cirz},~\bfnm{Ryan~T.}\binits{R.~T.}},
  \bauthor{\bsnm{Chin},~\bfnm{Jodie~K.}\binits{J.~K.}},
  \bauthor{\bsnm{Andes},~\bfnm{David~R.}\binits{D.~R.}},
  \bauthor{\bparticle{de}
  \bsnm{Cr{\'{e}}cy-Lagard},~\bfnm{Val{\'{e}}rie}\binits{V.}},
  \bauthor{\bsnm{Craig},~\bfnm{William~A.}\binits{W.~A.}} \AND
  \bauthor{\bsnm{Romesberg},~\bfnm{Floyd~E.}\binits{F.~E.}}
(\byear{2005}).
\btitle{Inhibition of mutation and combating the evolution of antibiotic
  resistance}.
\bjournal{PLoS Biol.}
\bvolume{3}
\bpages{e176}.
\bid{doi={10.1371/journal.pbio.0030176}, issn={1545-7885},
  pii={04-PLBI-RA-0918R2}, pmcid={1088971}, pmid={15869329}}
\bptok{imsref}%
\end{barticle}
\endbibitem

\bibitem[\protect\citeauthoryear{Conlon et~al.}{2003}]{Conetal03}
\begin{barticle}[pbm]
\bauthor{\bsnm{Conlon},~\bfnm{Erin~M.}\binits{E.~M.}},
  \bauthor{\bsnm{Liu},~\bfnm{X.~Shirley}\binits{X.~S.}},
  \bauthor{\bsnm{Lieb},~\bfnm{Jason~D.}\binits{J.~D.}} \AND
  \bauthor{\bsnm{Liu},~\bfnm{Jun~S.}\binits{J.~S.}}
(\byear{2003}).
\btitle{Integrating regulatory motif discovery and genome-wide expression
  analysis}.
\bjournal{Proc. Natl. Acad. Sci. USA}
\bvolume{100}
\bpages{3339--3344}.
\bid{doi={10.1073/pnas.0630591100}, issn={0027-8424}, pii={0630591100},
  pmcid={152294}, pmid={12626739}}
\bptok{imsref}%
\end{barticle}
\endbibitem

\bibitem[\protect\citeauthoryear{Courcelle et~al.}{2001}]{Couetal01}
\begin{barticle}[auto:STB|2011/09/12|07:03:23]
\bauthor{\bsnm{Courcelle},~\bfnm{J.}\binits{J.}},
  \bauthor{\bsnm{Khodursky},~\bfnm{A.}\binits{A.}},
  \bauthor{\bsnm{Peter},~\bfnm{B.}\binits{B.}},
  \bauthor{\bsnm{Brown},~\bfnm{P.~O.}\binits{P.~O.}} \AND
  \bauthor{\bsnm{Hanawalt},~\bfnm{P.~C.}\binits{P.~C.}}
(\byear{2001}).
\btitle{Comparative gene expression profiles following UV exposure in wild-type
  and SOS-deficient \textit{Escherichia coli}}.
\bjournal{Genetics}
\bvolume{158}
\bpages{41--64}.
\bptok{imsref}%
\end{barticle}
\endbibitem

\bibitem[\protect\citeauthoryear{Dempster, Laird and Rubin}{1977}]{DemLaiRub77}
\begin{barticle}[mr]
\bauthor{\bsnm{Dempster},~\bfnm{A.~P.}\binits{A.~P.}},
  \bauthor{\bsnm{Laird},~\bfnm{N.~M.}\binits{N.~M.}} \AND
  \bauthor{\bsnm{Rubin},~\bfnm{D.~B.}\binits{D.~B.}}
(\byear{1977}).
\btitle{Maximum likelihood from incomplete data via the {EM} algorithm}.
\bjournal{J. Roy. Statist. Soc. Ser. B}
\bvolume{39}
\bpages{1--38}.
\bid{issn={0035-9246}, mr={0501537}}
\bptnote{check related}%
\bptok{imsref}%
\end{barticle}
\endbibitem

\bibitem[\protect\citeauthoryear{Deng, Chen and Sun}{2004}]{DenCheSun04}
\begin{barticle}[auto:STB|2011/09/12|07:03:23]
\bauthor{\bsnm{Deng},~\bfnm{M.~H.}\binits{M.~H.}},
  \bauthor{\bsnm{Chen},~\bfnm{T.}\binits{T.}} \AND
  \bauthor{\bsnm{Sun},~\bfnm{F.}\binits{F.}}
(\byear{2004}).
\btitle{An integrated probabilistic model for functional prediction of
  proteins}.
\bjournal{J. Comput. Biol.}
\bvolume{11}
\bpages{463--475}.
\bptok{imsref}%
\end{barticle}
\endbibitem

\bibitem[\protect\citeauthoryear{Faith et~al.}{2008}]{Faietal}
\begin{barticle}[auto:STB|2011/09/12|07:03:23]
\bauthor{\bsnm{Faith},~\bfnm{J.~J.}\binits{J.~J.}},
  \bauthor{\bsnm{Driscoll},~\bfnm{M.~E.}\binits{M.~E.}},
  \bauthor{\bsnm{Fusaro},~\bfnm{V.~A.}\binits{V.~A.}},
  \bauthor{\bsnm{Cosgrove},~\bfnm{E.~J.}\binits{E.~J.}},
  \bauthor{\bsnm{Hayete},~\bfnm{B.}\binits{B.}},
  \bauthor{\bsnm{Juhn},~\bfnm{F.~S.}\binits{F.~S.}},
  \bauthor{\bsnm{Schneider},~\bfnm{S.~J.}\binits{S.~J.}} \AND
  \bauthor{\bsnm{Gardner},~\bfnm{T.~S.}\binits{T.~S.}}
(\byear{2008}).
\btitle{Many microbe microarrays database: Uniformly normalized
  Affymetrix compendia with structured experimental metadata}.
\bjournal{Nucleic Acids Res.}
\bvolume{36}
\bpages{D866--D870}.
\bptok{imsref}%
\end{barticle}
\endbibitem

\bibitem[\protect\citeauthoryear{Franke et~al.}{2006}]{Fraetal06}
\begin{barticle}[pbm]
\bauthor{\bsnm{Franke},~\bfnm{Lude}\binits{L.}}, \bauthor{\bparticle{van}
  \bsnm{Bakel},~\bfnm{Harm}\binits{H.}},
  \bauthor{\bsnm{Fokkens},~\bfnm{Like}\binits{L.}}, \bauthor{\bparticle{de}
  \bsnm{Jong},~\bfnm{Edwin~D.}\binits{E.~D.}},
  \bauthor{\bsnm{Egmont-Petersen},~\bfnm{Michael}\binits{M.}} \AND
  \bauthor{\bsnm{Wijmenga},~\bfnm{Cisca}\binits{C.}}
(\byear{2006}).
\btitle{Reconstruction of a functional human gene network, with an application
  for prioritizing positional candidate genes}.
\bjournal{Am. J. Hum. Genet.}
\bvolume{78}
\bpages{1011--1025}.
\bid{doi={10.1086/504300}, issn={0002-9297}, pii={S0002-9297(07)63922-6},
  pmcid={1474084}, pmid={16685651}}
\bptok{imsref}%
\end{barticle}
\endbibitem

\bibitem[\protect\citeauthoryear{Gama-Castro et~al.}{2008}]{Gametal08}
\begin{barticle}[pbm]
\bauthor{\bsnm{Gama-Castro},~\bfnm{Socorro}\binits{S.}},
  \bauthor{\bsnm{Jim{\'{e}}nez-Jacinto},~\bfnm{Ver{\'{o}}nica}\binits{V.}},
  \bauthor{\bsnm{Peralta-Gil},~\bfnm{Mart{\'{\i}}n}\binits{M.}},
  \bauthor{\bsnm{Santos-Zavaleta},~\bfnm{Alberto}\binits{A.}},
  \bauthor{\bsnm{Pe{\~{n}}aloza-Spinola},~\bfnm{M{\'{o}}nica~I.}\binits{M.~I.}%
}, \bauthor{\bsnm{Contreras-Moreira},~\bfnm{Bruno}\binits{B.}},
  \bauthor{\bsnm{Segura-Salazar},~\bfnm{Juan}\binits{J.}},
  \bauthor{\bsnm{Mu{\~{n}}iz-Rascado},~\bfnm{Luis}\binits{L.}},
  \bauthor{\bsnm{Mart{\'{\i}}nez-Flores},~\bfnm{Irma}\binits{I.}},
  \bauthor{\bsnm{Salgado},~\bfnm{Heladia}\binits{H.}},
  \bauthor{\bsnm{Bonavides-Mart{\'{\i}}nez},~\bfnm{C{\'{e}}sar}\binits{C.}},
  \bauthor{\bsnm{Abreu-Goodger},~\bfnm{Cei}\binits{C.}},
  \bauthor{\bsnm{Rodr{\'{\i}}guez-Penagos},~\bfnm{Carlos}\binits{C.}},
  \bauthor{\bsnm{Miranda-R{\'{\i}}os},~\bfnm{Juan}\binits{J.}},
  \bauthor{\bsnm{Morett},~\bfnm{Enrique}\binits{E.}},
  \bauthor{\bsnm{Merino},~\bfnm{Enrique}\binits{E.}},
  \bauthor{\bsnm{Huerta},~\bfnm{Araceli~M.}\binits{A.~M.}},
  \bauthor{\bsnm{Trevi{\~{n}}o-Quintanilla},~\bfnm{Luis}\binits{L.}} \AND
  \bauthor{\bsnm{Collado-Vides},~\bfnm{Julio}\binits{J.}}
(\byear{2008}).
\btitle{RegulonDB (version 6.0): Gene regulation model of Escherichia coli K-12
  beyond transcription, active (experimental) annotated promoters and
  Textpresso navigation}.
\bjournal{Nucleic Acids Res.}
\bvolume{36}
\bpages{D120--D124}.
\bid{doi={10.1093/nar/gkm994}, issn={1362-4962}, pii={gkm994}, pmcid={2238961},
  pmid={18158297}}
\bptok{imsref}%
\end{barticle}
\endbibitem

\bibitem[\protect\citeauthoryear{Gelman and Rubin}{1992}]{Dav92}
\begin{barticle}[auto]
\bauthor{\bsnm{Gelman},~\bfnm{A.}\binits{A.}} \AND
  \bauthor{\bsnm{Rubin},~\bfnm{D.~B.}\binits{D.~B.}}
(\byear{1992}).
\btitle{Inference from iterative simulation using multiple sequences (with
  discussion)}.
\bjournal{Statist. Sci.}
\bvolume{7}
\bpages{457--511}.
\bptok{imsref}%
\end{barticle}
\endbibitem

\bibitem[\protect\citeauthoryear{Jensen, Chen and
  Stoeckert}{2007}]{JenCheSto07}
\begin{barticle}[mr]
\bauthor{\bsnm{Jensen},~\bfnm{Shane~T.}\binits{S.~T.}},
  \bauthor{\bsnm{Chen},~\bfnm{Guang}\binits{G.}} \AND
  \bauthor{\bsnm{Stoeckert},~\bfnm{Christian~J.}\binits{C.~J.} \bsuffix{Jr.}}
(\byear{2007}).
\btitle{Bayesian variable selection and data integration for biological
  regulatory networks}.
\bjournal{Ann. Appl. Stat.}
\bvolume{1}
\bpages{612--633}.
\bid{doi={10.1214/07-AOAS130}, issn={1932-6157}, mr={2415749}}
\bptok{imsref}%
\end{barticle}
\endbibitem

\bibitem[\protect\citeauthoryear{Kanehisa and Goto}{2002}]{KanGot02}
\begin{barticle}[auto:STB|2011/09/12|07:03:23]
\bauthor{\bsnm{Kanehisa},~\bfnm{M.}\binits{M.}} \AND
  \bauthor{\bsnm{Goto},~\bfnm{S.}\binits{S.}}
(\byear{2002}).
\btitle{KEGG: Kyoto encyclopedia of genes and genomes}.
\bjournal{Nucleic Acids Res.}
\bvolume{28}
\bpages{27--30}.
\bptok{imsref}%
\end{barticle}
\endbibitem

\bibitem[\protect\citeauthoryear{Li and Li}{2008}]{LiLi08}
\begin{barticle}[auto:STB|2011/09/12|07:03:23]
\bauthor{\bsnm{Li},~\bfnm{C.}\binits{C.}} \AND
  \bauthor{\bsnm{Li},~\bfnm{H.}\binits{H.}}
(\byear{2008}).
\btitle{Network-constrained regularization and variable selection for analysis
  of genomic data}.
\bjournal{Bioinformatics}
\bvolume{24}
\bpages{1175--1182}.
\bptok{imsref}%
\end{barticle}
\endbibitem

\bibitem[\protect\citeauthoryear{Michel}{2005}]{Mic05}
\begin{barticle}[pbm]
\bauthor{\bsnm{Michel},~\bfnm{B{\'{e}}n{\'{e}}dicte}\binits{B.}}
(\byear{2005}).
\btitle{After 30 years of study, the bacterial SOS response still surprises
  us}.
\bjournal{PLoS Biol.}
\bvolume{3}
\bpages{e255}.
\bid{doi={10.1371/journal.pbio.0030255}, issn={1545-7885},
  pii={05-PLBI-P-0386}, pmcid={1174825}, pmid={16000023}}
\bptok{imsref}%
\end{barticle}
\endbibitem

\bibitem[\protect\citeauthoryear{M{\o}ller et~al.}{2006}]{Mlletal06}
\begin{barticle}[mr]
\bauthor{\bsnm{M{\o}ller},~\bfnm{J.}\binits{J.}},
  \bauthor{\bsnm{Pettitt},~\bfnm{A.~N.}\binits{A.~N.}},
  \bauthor{\bsnm{Reeves},~\bfnm{R.}\binits{R.}} \AND
  \bauthor{\bsnm{Berthelsen},~\bfnm{K.~K.}\binits{K.~K.}}
(\byear{2006}).
\btitle{An efficient {M}arkov chain {M}onte {C}arlo method for distributions
  with intractable normalising constants}.
\bjournal{Biometrika}
\bvolume{93}
\bpages{451--458}.
\bid{doi={10.1093/biomet/93.2.451}, issn={0006-3444}, mr={2278096}}
\bptok{imsref}%
\end{barticle}
\endbibitem

\bibitem[\protect\citeauthoryear{Pan, Wei and Khodursky}{2008}]{PanWeiKho08}
\begin{barticle}[pbm]
\bauthor{\bsnm{Pan},~\bfnm{Wei}\binits{W.}},
  \bauthor{\bsnm{Wei},~\bfnm{Peng}\binits{P.}} \AND
  \bauthor{\bsnm{Khodursky},~\bfnm{Arkady}\binits{A.}}
(\byear{2008}).
\btitle{A parametric joint model of DNA-protein binding, gene expression and
  DNA sequence data to detect target genes of a transcription factor}.
\bjournal{Pac. Symp. Biocomput.}
\bvolume{13}
\bpages{465--476}.
\bid{issn={1793-5091}, pmid={18229708}}
\bptok{imsref}%
\end{barticle}
\endbibitem

\bibitem[\protect\citeauthoryear{Prasad et~al.}{2009}]{Praetal09}
\begin{barticle}[pbm]
\bauthor{\bsnm{Prasad},~\bfnm{T.~S.~Keshava}\binits{T.~S.~K.}},
  \bauthor{\bsnm{Goel},~\bfnm{Renu}\binits{R.}},
  \bauthor{\bsnm{Kandasamy},~\bfnm{Kumaran}\binits{K.}},
  \bauthor{\bsnm{Keerthikumar},~\bfnm{Shivakumar}\binits{S.}},
  \bauthor{\bsnm{Kumar},~\bfnm{Sameer}\binits{S.}},
  \bauthor{\bsnm{Mathivanan},~\bfnm{Suresh}\binits{S.}},
  \bauthor{\bsnm{Telikicherla},~\bfnm{Deepthi}\binits{D.}},
  \bauthor{\bsnm{Raju},~\bfnm{Rajesh}\binits{R.}},
  \bauthor{\bsnm{Shafreen},~\bfnm{Beema}\binits{B.}},
  \bauthor{\bsnm{Venugopal},~\bfnm{Abhilash}\binits{A.}},
  \bauthor{\bsnm{Balakrishnan},~\bfnm{Lavanya}\binits{L.}},
  \bauthor{\bsnm{Marimuthu},~\bfnm{Arivusudar}\binits{A.}},
  \bauthor{\bsnm{Banerjee},~\bfnm{Sutopa}\binits{S.}},
  \bauthor{\bsnm{Somanathan},~\bfnm{Devi~S.}\binits{D.~S.}},
  \bauthor{\bsnm{Sebastian},~\bfnm{Aimy}\binits{A.}},
  \bauthor{\bsnm{Rani},~\bfnm{Sandhya}\binits{S.}},
  \bauthor{\bsnm{Ray},~\bfnm{Somak}\binits{S.}},
  \bauthor{\bsnm{Kishore},~\bfnm{C.~J.~Harrys}\binits{C.~J.~H.}},
  \bauthor{\bsnm{Kanth},~\bfnm{Sashi}\binits{S.}},
  \bauthor{\bsnm{Ahmed},~\bfnm{Mukhtar}\binits{M.}},
  \bauthor{\bsnm{Kashyap},~\bfnm{Manoj~K.}\binits{M.~K.}},
  \bauthor{\bsnm{Mohmood},~\bfnm{Riaz}\binits{R.}},
  \bauthor{\bsnm{Ramachandra},~\bfnm{Y.~L.}\binits{Y.~L.}},
  \bauthor{\bsnm{Krishna},~\bfnm{V.}\binits{V.}},
  \bauthor{\bsnm{Rahiman},~\bfnm{B.~Abdul}\binits{B.~A.}},
  \bauthor{\bsnm{Mohan},~\bfnm{Sujatha}\binits{S.}},
  \bauthor{\bsnm{Ranganathan},~\bfnm{Prathibha}\binits{P.}},
  \bauthor{\bsnm{Ramabadran},~\bfnm{Subhashri}\binits{S.}},
  \bauthor{\bsnm{Chaerkady},~\bfnm{Raghothama}\binits{R.}} \AND
  \bauthor{\bsnm{Pandey},~\bfnm{Akhilesh}\binits{A.}}
(\byear{2009}).
\btitle{Human protein reference database--2009 update}.
\bjournal{Nucleic Acids Res.}
\bvolume{37}
\bpages{D767--D772}.
\bid{doi={10.1093/nar/gkn892}, issn={1362-4962}, pii={gkn892}, pmcid={2686490},
  pmid={18988627}}
\bptnote{check year}%
\bptok{imsref}%
\end{barticle}
\endbibitem

\bibitem[\protect\citeauthoryear{Roth et~al.}{1998}]{Rotetal98}
\begin{barticle}[auto:STB|2011/09/12|07:03:23]
\bauthor{\bsnm{Roth},~\bfnm{F.~P.}\binits{F.~P.}},
  \bauthor{\bsnm{Hughes},~\bfnm{J.~D.}\binits{J.~D.}},
  \bauthor{\bsnm{Estep},~\bfnm{P.~W.}\binits{P.~W.}} \AND
  \bauthor{\bsnm{Church},~\bfnm{G.~M.}\binits{G.~M.}}
(\byear{1998}).
\btitle{Finding DNA regulatory motifs within unaligned noncoding sequences
  clustered by whole-genome mRNA quantitation}.
\bjournal{Nat. Biotech.}
\bvolume{16}
\bpages{939--945}.
\bptok{imsref}%
\end{barticle}
\endbibitem

\bibitem[\protect\citeauthoryear{Ryd{\'e}n and Titterington}{1998}]{RydTit98}
\begin{barticle}[mr]
\bauthor{\bsnm{Ryd{\'e}n},~\bfnm{Tobias}\binits{T.}} \AND
  \bauthor{\bsnm{Titterington},~\bfnm{D.~M.}\binits{D.~M.}}
(\byear{1998}).
\btitle{Computational {B}ayesian analysis of hidden {M}arkov models}.
\bjournal{J. Comput. Graph. Statist.}
\bvolume{7}
\bpages{194--211}.
\bid{doi={10.2307/1390813}, issn={1061-8600}, mr={1649362}}
\bptok{imsref}%
\end{barticle}
\endbibitem

\bibitem[\protect\citeauthoryear{Spiegelhalter et~al.}{2003}]{Spietal03}
\begin{bmisc}[auto:STB|2011/09/12|07:03:23]
\bauthor{\bsnm{Spiegelhalter},~\bfnm{D.}\binits{D.}},
  \bauthor{\bsnm{Thomas},~\bfnm{A.}\binits{A.}},
  \bauthor{\bsnm{Best},~\bfnm{N.}\binits{N.}} \AND
  \bauthor{\bsnm{Lunn},~\bfnm{D.}\binits{D.}}
(\byear{2003}).
\bhowpublished{WinBUGS User Manual, Version 1.4. Available at
\texttt{\href{http://www.mrc-bsu.cam.ac.uk/bugs/winbugs/manual14.pdf}{http://www.mrc-bsu.cam.ac.uk/bugs/winbugs/}
\href{http://www.mrc-bsu.cam.ac.uk/bugs/winbugs/manual14.pdf}{manual14.pdf}}.}
\bptok{imsref}%
\end{bmisc}
\endbibitem

\bibitem[\protect\citeauthoryear{Sun, Carroll and Zhao}{2006}]{SunCarZha06}
\begin{barticle}[pbm]
\bauthor{\bsnm{Sun},~\bfnm{Ning}\binits{N.}},
  \bauthor{\bsnm{Carroll},~\bfnm{Raymond~J.}\binits{R.~J.}} \AND
  \bauthor{\bsnm{Zhao},~\bfnm{Hongyu}\binits{H.}}
(\byear{2006}).
\btitle{Bayesian error analysis model for reconstructing transcriptional
  regulatory networks}.
\bjournal{Proc. Natl. Acad. Sci. USA}
\bvolume{103}
\bpages{7988--7993}.
\bid{doi={10.1073/pnas.0600164103}, issn={0027-8424}, pii={0600164103},
  pmcid={1472417}, pmid={16702552}}
\bptok{imsref}%
\end{barticle}
\endbibitem

\bibitem[\protect\citeauthoryear{Wade et~al.}{2005}]{Wadetal05}
\begin{barticle}[pbm]
\bauthor{\bsnm{Wade},~\bfnm{Joseph~T.}\binits{J.~T.}},
  \bauthor{\bsnm{Reppas},~\bfnm{Nikos~B.}\binits{N.~B.}},
  \bauthor{\bsnm{Church},~\bfnm{George~M.}\binits{G.~M.}} \AND
  \bauthor{\bsnm{Struhl},~\bfnm{Kevin}\binits{K.}}
(\byear{2005}).
\btitle{Genomic analysis of LexA binding reveals the permissive nature of the
  \textit{Escherichia coli} genome and identifies unconventional target sites}.
\bjournal{Genes Dev.}
\bvolume{19}
\bpages{2619--2630}.
\bid{doi={10.1101/gad.1355605}, issn={0890-9369}, pii={19/21/2619},
  pmcid={1276735}, pmid={16264194}}
\bptok{imsref}%
\end{barticle}
\endbibitem

\bibitem[\protect\citeauthoryear{Wang et~al.}{2005}]{Wanetal05}
\begin{barticle}[pbm]
\bauthor{\bsnm{Wang},~\bfnm{Wei}\binits{W.}},
  \bauthor{\bsnm{Cherry},~\bfnm{J.~Michael}\binits{J.~M.}},
  \bauthor{\bsnm{Nochomovitz},~\bfnm{Yigal}\binits{Y.}},
  \bauthor{\bsnm{Jolly},~\bfnm{Emmitt}\binits{E.}},
  \bauthor{\bsnm{Botstein},~\bfnm{David}\binits{D.}} \AND
  \bauthor{\bsnm{Li},~\bfnm{Hao}\binits{H.}}
(\byear{2005}).
\btitle{Inference of combinatorial regulation in yeast transcriptional
  networks: A case study of sporulation}.
\bjournal{Proc. Natl. Acad. Sci. USA}
\bvolume{102}
\bpages{1998--2003}.
\bid{doi={10.1073/pnas.0405537102}, issn={0027-8424}, pii={0405537102},
  pmcid={548531}, pmid={15684073}}
\bptok{imsref}%
\end{barticle}
\endbibitem

\bibitem[\protect\citeauthoryear{Wei and Li}{2007}]{WeiLi07}
\begin{barticle}[pbm]
\bauthor{\bsnm{Wei},~\bfnm{Zhi}\binits{Z.}} \AND
  \bauthor{\bsnm{Li},~\bfnm{Hongzhe}\binits{H.}}
(\byear{2007}).
\btitle{A Markov random field model for network-based analysis of genomic
  data}.
\bjournal{Bioinformatics}
\bvolume{23}
\bpages{1537--1544}.
\bid{doi={10.1093/bioinformatics/btm129}, issn={1367-4811}, pii={btm129},
  pmid={17483504}}
\bptok{imsref}%
\end{barticle}
\endbibitem

\bibitem[\protect\citeauthoryear{Wei and Li}{2008}]{WeiLi08}
\begin{barticle}[mr]
\bauthor{\bsnm{Wei},~\bfnm{Zhi}\binits{Z.}} \AND
  \bauthor{\bsnm{Li},~\bfnm{Hongzhe}\binits{H.}}
(\byear{2008}).
\btitle{A hidden spatial-temporal {M}arkov random field model for network-based
  analysis of time course gene expression data}.
\bjournal{Ann. Appl. Stat.}
\bvolume{2}
\bpages{408--429}.
\bid{doi={10.1214/07--AOAS145}, issn={1932-6157}, mr={2415609}}
\bptok{imsref}%
\end{barticle}
\endbibitem

\bibitem[\protect\citeauthoryear{Wei and Pan}{2008a}]{WeiPan08N1}
\begin{barticle}[auto:STB|2011/09/12|07:03:23]
\bauthor{\bsnm{Wei},~\bfnm{P.}\binits{P.}} \AND
  \bauthor{\bsnm{Pan},~\bfnm{W.}\binits{W.}}
(\byear{2008}a).
\btitle{Incorporating gene networks into statistical tests for genomic data via
  a spatially correlated mixture model}.
\bjournal{Bioinformatics}
\bvolume{24}
\bpages{404--411}.
\bptok{imsref}%
\end{barticle}
\endbibitem

\bibitem[\protect\citeauthoryear{Wei and Pan}{2008b}]{WeiPan08N2}
\begin{barticle}[auto:STB|2011/09/12|07:03:23]
\bauthor{\bsnm{Wei},~\bfnm{P.}\binits{P.}} \AND
  \bauthor{\bsnm{Pan},~\bfnm{W.}\binits{W.}}
(\byear{2008}b).
\btitle{Incorporating gene functions into regression analysis of DNA-protein
  binding data and gene expression data to construct transcriptional networks}.
\bjournal{IEEE/ACM Transactions on Computational Biology and Bioinformatics}
\bvolume{5}
\bpages{401--415}.
\bptok{imsref}%
\end{barticle}
\endbibitem

\bibitem[\protect\citeauthoryear{Wei and Pan}{2010}]{WeiPan10}
\begin{barticle}[mr]
\bauthor{\bsnm{Wei},~\bfnm{Peng}\binits{P.}} \AND
  \bauthor{\bsnm{Pan},~\bfnm{Wei}\binits{W.}}
(\byear{2010}).
\btitle{Network-based genomic discovery: Application and comparison of {M}arkov
  random-field models}.
\bjournal{J. R. Stat. Soc. Ser. C Appl. Stat.}
\bvolume{59}
\bpages{105--125}.
\bid{doi={10.1111/j.1467-9876.2009.00686.x}, issn={0035-9254}, mr={2750134}}
\bptok{imsref}%
\end{barticle}
\endbibitem

\bibitem[\protect\citeauthoryear{Wei and Pan}{2011}]{WeiPan}
\begin{bmisc}[auto:STB|2011/09/12|07:03:23]
\bauthor{\bsnm{Wei},~\bfnm{P.}\binits{P.}} \AND
  \bauthor{\bsnm{Pan},~\bfnm{W.}\binits{W.}}
(\byear{2011}).
\bhowpublished{Supplement to ``Bayesian joint modeling of multiple gene
  networks and diverse genomic data to identify target genes of a transcription
  factor.''
\href{http://dx.doi.org/10.1214/11-AOAS502SUPP}{DOI:10.1214/11-AOAS502SUPP}.}
\bptok{imsref}%
\end{bmisc}
\endbibitem

\bibitem[\protect\citeauthoryear{Winkler}{2003}]{Win03}
\begin{bbook}[mr]
\bauthor{\bsnm{Winkler},~\bfnm{Gerhard}\binits{G.}}
(\byear{2003}).
\btitle{Image Analysis, Random Fields and {M}arkov Chain {M}onte {C}arlo
  Methods: A Mathematical Introduction},
\bedition{2nd} ed.
\bseries{Applications of Mathematics (New York)}
\bvolume{27}.
\bpublisher{Springer}, \baddress{Berlin}.
\bid{mr={1950762}}
\bptnote{check year}%
\bptok{imsref}%
\end{bbook}
\endbibitem

\bibitem[\protect\citeauthoryear{Wu et~al.}{2005}]{Wuetal05}
\begin{barticle}[pbm]
\bauthor{\bsnm{Wu},~\bfnm{Hongwei}\binits{H.}},
  \bauthor{\bsnm{Su},~\bfnm{Zhengchang}\binits{Z.}},
  \bauthor{\bsnm{Mao},~\bfnm{Fenglou}\binits{F.}},
  \bauthor{\bsnm{Olman},~\bfnm{Victor}\binits{V.}} \AND
  \bauthor{\bsnm{Xu},~\bfnm{Ying}\binits{Y.}}
(\byear{2005}).
\btitle{Prediction of functional modules based on comparative genome analysis
  and gene ontology application}.
\bjournal{Nucleic Acids Res.}
\bvolume{33}
\bpages{2822--2837}.
\bid{doi={10.1093/nar/gki573}, issn={1362-4962}, pii={33/9/2822},
  pmcid={1130488}, pmid={15901854}}
\bptok{imsref}%
\end{barticle}
\endbibitem

\bibitem[\protect\citeauthoryear{Xie et~al.}{2010}]{Xieetal10}
\begin{barticle}[mr]
\bauthor{\bsnm{Xie},~\bfnm{Yang}\binits{Y.}},
  \bauthor{\bsnm{Pan},~\bfnm{Wei}\binits{W.}},
  \bauthor{\bsnm{Jeong},~\bfnm{Kyeong~S.}\binits{K.~S.}},
  \bauthor{\bsnm{Xiao},~\bfnm{Guanghua}\binits{G.}} \AND
  \bauthor{\bsnm{Khodursky},~\bfnm{Arkady~B.}\binits{A.~B.}}
(\byear{2010}).
\btitle{A {B}ayesian approach to joint modeling of protein-{DNA} binding, gene
  expression and sequence data}.
\bjournal{Stat. Med.}
\bvolume{29}
\bpages{489--503}.
\bid{issn={0277-6715}, mr={2751784}}
\bptok{imsref}%
\end{barticle}
\endbibitem

\bibitem[\protect\citeauthoryear{Yang and Dudoit}{2002}]{YanDud02}
\begin{barticle}[auto:STB|2011/09/12|07:03:23]
\bauthor{\bsnm{Yang},~\bfnm{Y.~H.}\binits{Y.~H.}} \AND
  \bauthor{\bsnm{Dudoit},~\bfnm{et~al.}\binits{e.~a.} \bsuffix{S.}}
(\byear{2002}).
\btitle{Normalization for cDNA microarray data: A robust composite method
  addressing single and multiple slide systematic variation}.
\bjournal{Nucleic Acids Res.}
\bvolume{304}
\bpages{e15}.
\bptok{imsref}%
\end{barticle}
\endbibitem

\bibitem[\protect\citeauthoryear{Zhang, Pigli and Rice}{2010}]{ZhaPigRic10}
\begin{barticle}[auto:STB|2011/09/12|07:03:23]
\bauthor{\bsnm{Zhang},~\bfnm{A.~P.~P.}\binits{A.~P.~P.}},
  \bauthor{\bsnm{Pigli},~\bfnm{Y.~Z.}\binits{Y.~Z.}} \AND
  \bauthor{\bsnm{Rice},~\bfnm{P.~A.}\binits{P.~A.}}
(\byear{2010}).
\btitle{Structure of the LexA-DNA complex and implications for SOS box
  measurement}.
\bjournal{Nature}
\bvolume{466}
\bpages{883--886}.
\bptok{imsref}%
\end{barticle}
\endbibitem

\end{thebibliography}
\end{document}